\documentclass{article}
\usepackage{amssymb,graphicx,cancel}
\usepackage[margin=2cm]{geometry}
\usepackage{slashed}
\usepackage{hepunits}
\usepackage{color}
\usepackage[table]{xcolor}
\usepackage{graphicx}

 \usepackage{jheppub}
\begin{document}
\newcommand{\be}{\begin{eqnarray}}
\newcommand{\ee}{\end{eqnarray}}
\newcommand{\Gev}{\,\,\mathrm{GeV}}
\newcommand{\SUWeak}{\mathrm{SU}(2)_{\rm W}}
\newcommand{\Lag}{\mathcal{L}}
\newcommand{\Lagtree}{\mathcal{L}_{\rm tree}}
\newcommand{\benum}{\begin{enumerate}}
\newcommand{\eenum}{\end{enumerate}}
\newcommand{\bi}{\begin{itemize}}
\newcommand{\ei}{\end{itemize}}
\newcommand{\met}{\slashed{E_T}}
\newcommand{\apr}{{A^\prime}}
\newcommand{\zp}{{Z^\prime}}
\newcommand{\zpr}{{Z^\prime}}

\preprint{\\ FERMILAB-PUB-18-087-A, PUPT-2557}

\newcommand\FNAL{Fermi National Accelerator Laboratory, \\ Batavia, IL USA}
\newcommand\Princeton{Princeton University, \\ Princeton, NJ USA}
\newcommand\UCI{University of California, Irvine, \\ Irvine, CA USA}

\newcommand{\gk}[1]{\textbf{\textcolor{red}{(#1 --gk)}}}
\newcommand{\yk}[1]{\textbf{\textcolor{blue}{(#1 --yk)}}}
\newcommand{\nt}[1]{\textbf{\textcolor{orange}{(#1 --nt)}}}
\newcommand{\aw}[1]{\textbf{\textcolor{green}{(#1 --aw)}}}

\author[a]{Yonatan Kahn,} 
\author[b]{Gordan Krnjaic,}
\author[b]{Nhan Tran,}
\author[b]{Andrew Whitbeck}

\affiliation[a]{\Princeton}
\affiliation[b]{\FNAL}

\abstract{
New light, weakly-coupled particles are commonly invoked to address the persistent $\sim 4\sigma$ anomaly in $(g-2)_\mu$ and serve as mediators between dark and
visible matter. If such particles couple predominantly to heavier generations and decay invisibly, much of their best-motivated parameter 
space is inaccessible with existing experimental techniques.
 In this paper, we present a new fixed-target, missing-momentum search strategy to probe invisibly decaying particles that couple preferentially to muons. 
In our setup, a relativistic muon beam  impinges on a thick active target. 
The signal consists of events in which a muon loses a large fraction of its incident momentum inside the target
 without initiating any detectable electromagnetic or hadronic activity in downstream veto systems. 
We propose a two-phase experiment, M$^3$ (Muon Missing Momentum), based at Fermilab. Phase 1 with $\sim 10^{10}$ muons on target 
can test the remaining parameter space for which light invisibly-decaying particles can resolve the $(g-2)_\mu$ anomaly, while Phase 2 with $\sim 10^{13}$ muons on target can test much of the predictive parameter space over which
sub-GeV dark matter achieves freeze-out via muon-philic forces, including gauged 
$U(1)_{L_\mu - L_\tau}$. 
}

\title{ 
M$^3$: A New Muon Missing Momentum Experiment to Probe $(g-2)_{\mu}$ and
Dark Matter at Fermilab 
}
\maketitle

\section{Introduction}

Despite decades of dedicated searches, the particle nature of dark matter (DM) remains one of the greatest mysteries in all of physics (for a review see \cite{Bertone:2016nfn}).  
In response to null results from direct detection and collider experiments designed to probe weakly interacting massive particles, there has recently been a surge of activity in designing new techniques to explore ``hidden sectors" in which DM is a Standard Model (SM) singlet with its own dynamics and cosmological history (see \cite{Alexander:2016aln,Battaglieri:2017aum} for a review). In a well-motivated class of  such models, DM interacts predominantly with muons through a new force carrier, which enables thermal freeze-out in the early universe.  This scenario generally features sizable couplings to visible particles and predictive experimental benchmarks, 
yet remains inaccessible to traditional DM search strategies, thereby motivating new techniques. 

Such muon-specific forces are also well-motivated independently of any possible connection to dark matter. The 
 persistent $\sim 4 \sigma$ discrepancy in the anomalous magnetic moment of the muon remains one of the 
largest anomalies in particle physics \cite{Patrignani:2016xqp}.  It is well-known that light, weakly coupled particles can 
bring theoretical predictions into agreement with observations \cite{Pospelov:2008zw,Altmannshofer:2014pba}.  
However in recent years, the most popular new-physics explanation, a light ($<$ GeV)  dark photon with kinetic mixing $\epsilon \sim 10^{-3}$ and flavor-universal couplings, has been ruled out  
regardless of whether it decays visibly or invisibly \cite{Battaglieri:2017aum}. The only remaining class of light new-physics explanations involves  particles that couple predominantly (or exclusively) to muons, thereby evading the searches upon which the dark photon
exclusions are based.\footnote{New physics explanations with heavier electroweak states also remain viable, but are 
under tension from null LHC results -- see e.g. \cite{Endo:2013bba,Moroi:1995yh}.} Thus, a robust test of the new-physics hypothesis requires improved sensitivity to muonic forces.

\begin{figure}[t!] 
\hspace{-1.5cm}
\center
\includegraphics[width=7.75cm]{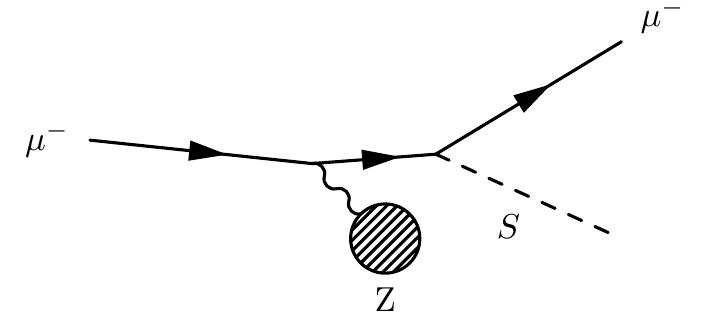} 
\hspace{-0.2cm}\includegraphics[width=7.75cm]{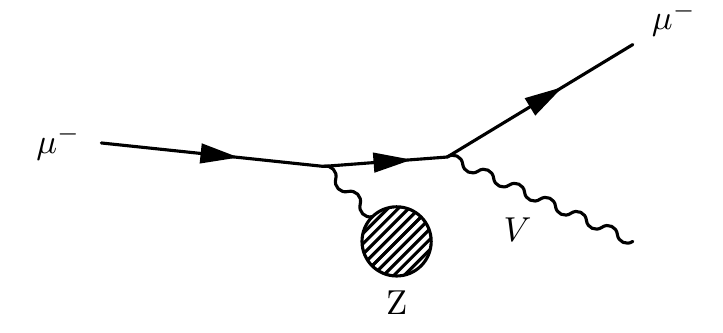} 
\caption{\label{fig:feyn} 
Dark bremsstrahlung signal process for simplified models with invisibly decaying 
scalar (\emph{left}) and vector (\emph{right}) forces that couple predominantly to muons. In both 
cases, a relativistic muon beam is incident on a fixed target and scatters 
coherently off a nucleus to produce the new particle as initial- or final-state radiation.
}
\vspace{0cm}
\end{figure}

One promising approach to probing invisibly decaying muonic forces is the NA64 experiment, which aims 
to utilize the CERN SPS muon beam to perform a missing energy search \cite{Gninenko:2014pea}. The technique exploits the radiative production process $\mu N \to \mu N \displaystyle{\not}{E}$  shown in Fig.~\ref{fig:feyn}, where the final state missing energy $\displaystyle{\not}{E}$ arises from the production of a new muon-philic particle ($S$ or $V$ for scalar or vector, respectively). Here the signal is defined by the kinematics of 
 the outgoing muon, which loses $\sim 50\%$ of its incident energy and
 momentum in a typical production event. Since the new radiated particle decays invisibly to DM or neutrinos, there is no additional visible energy, so
events with additional activity in an ECAL/HCAL downstream of the target are vetoed as SM-induced backgrounds. The projected sensitivity for this approach can cover the entire region of parameter space over which a muonic force can reconcile the $(g-2)_\mu$ anomaly for invisibly decaying, muon-philic bosons \cite{Gninenko:2014pea}. This approach complements the efforts of the
NA64 \cite{Banerjee:2016tad,Gninenko:2016kpg} and LDMX \cite{Izaguirre:2014bca,Mans:2017vej} experiments which are based on a similar setup, but utilize electron beams to probe invisibly-decaying particles which couple to electrons. The missing energy technique also complements proposed muon beam dump searches for {\it visibly decaying} muon-philic scalars proposed in \cite{Chen:2017awl}, which inherit couplings to electron and photons through mixing with the Higgs, as well as searches at BaBar for mediators with mass greater than $2m_\mu$ which can decay back to muons \cite{TheBABAR:2016rlg}.

In this paper, we extend the discussion in \cite{Gninenko:2014pea} to propose a new
fixed-target missing-momentum muon-beam experiment at Fermilab, which we dub Muon Missing Momentum or M$^3$, suitable for any muon-philic invisibly decaying particles. We also present thermal DM targets that such an effort could reach. Our study considers
the new physics discovery potential for a $15$~\GeV\ muon beam, which is representative of the Fermilab muon beam capabilities, paired with a thick ($\sim 50$ radiation length) target and a detector similar to the planned LDMX detector \cite{Mans:2017vej} downstream to veto SM backgrounds. The thicker active target compared to the $\sim 0.1$ radiation length in the nominal LDMX setup allows for a larger signal production rate while exploiting the fact that the muons will lose much less energy than electrons in a similarly-sized target. In analogy with similar processes involving electron beams, one can take advantage of the distinctive kinematics of the radiated massive scalar or vector particle $S,V$ to distinguish signal from background (see Fig.~\ref{fig:feyn}). The Fermilab muon beam option provides several advantages over existing proposals for new physics searches with either electron beams or high-energy muon beams:
\begin{itemize}
\item \textbf{Bremsstrahlung backgrounds suppressed.} The principal reducible backgrounds for LDMX are dominated by hadronic processes initiated by a real bremsstrahlung photon. Relative to electron beams, the M$^3$ bremsstrahlung rate is suppressed by $(m_e/m_\mu)^2 \approx 2 \times 10^{-5}$, so background rejection becomes much simpler for muon beams for an equivalent target thickness.

\item \textbf{Compact experimental design.} For $m_{S,V} \ll E_{\rm beam}$, the signal production cross section is largely independent of beam energy. However,
 compared to the CERN/SPS option \cite{Gninenko:2014pea}, with $\sim 100-200$ GeV beam muons, a lower-energy, e.g. $15$~\GeV, muon beam allows for greater muon track 
 curvature and, therefore, a more compact experimental design. In particular, percent-level momentum resolution is possible in M$^3$ with the target placed in the magnetic field region, reducing acceptance losses from having the magnet downstream of the target.
\end{itemize}

\begin{figure}[t!] 
\vspace{-0.2cm}
\center
\includegraphics[width=12cm]{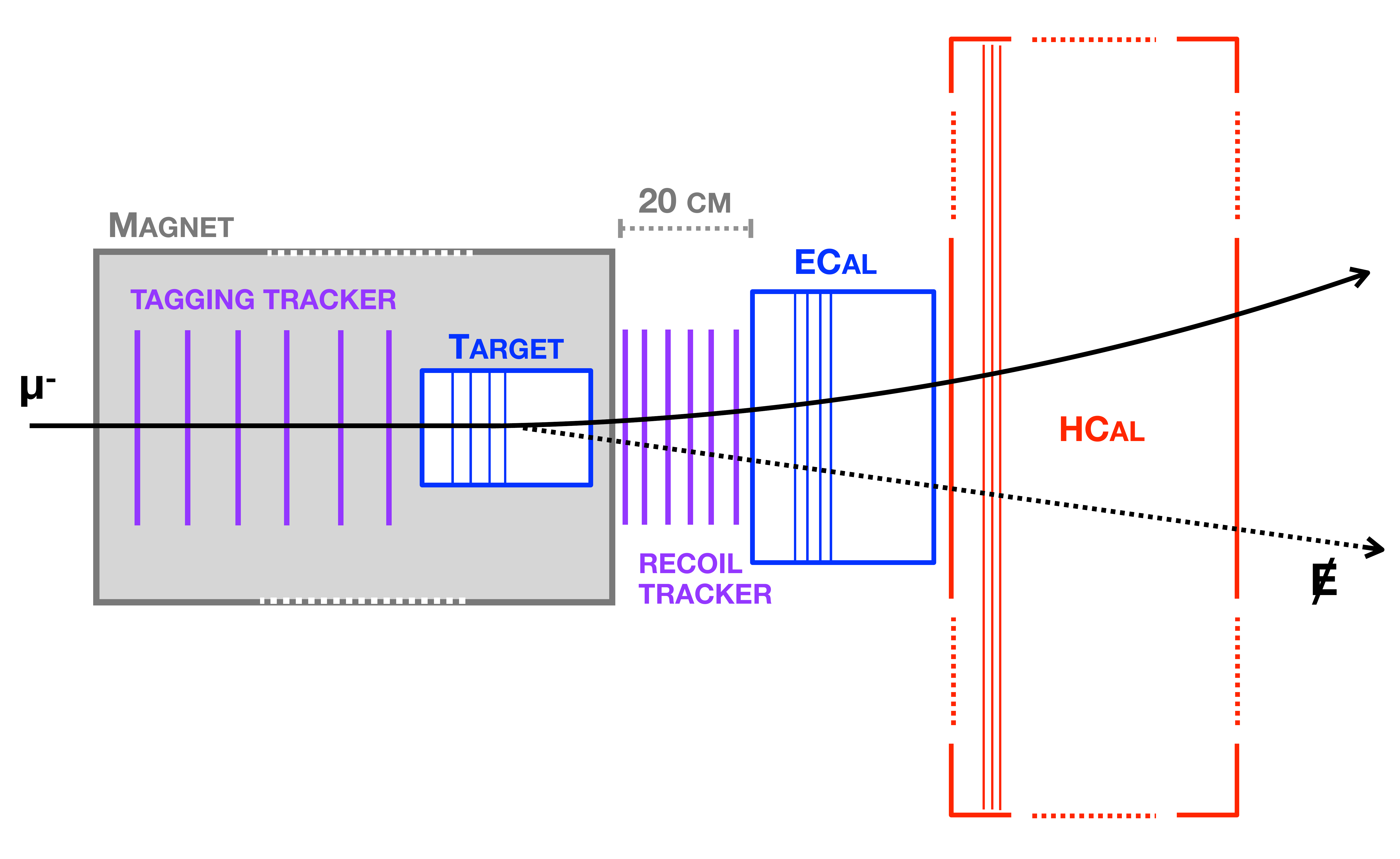} 
\caption{\label{fig:experiment} 
Experimental schematic. The incoming muon beam passes through a tagging tracker in the magnetic field region before entering the tungsten target. Outgoing muons are detected with a recoil tracker, with the magnet fringe field providing a momentum measurement. Electromagnetic and hadronic calorimeters veto on photons and hadrons produced in hard interactions in the target which could lead to significant muon energy loss.}
\vspace{0cm}
\end{figure}

\noindent We propose a two-phase experiment, each covering a well-motivated region of parameter space:

\begin{itemize}
\item \textbf{Phase 1:} $\mathbf{(g-2)_\mu}$ \textbf{search}. With $10^{10}$ muons on target (MOT) and existing detector technology, we will show that our setup can probe the entire $(g-2)_\mu$ region not currently excluded by experiments, for vectors with $m_V \lesssim 500 \ \MeV$ and scalars with $m_S \lesssim 100 \ \MeV$ which couple exclusively to muons and decay invisibly.\footnote{Models with a more complicated dark sector can fail our search criteria, for example an inelastic DM model $V \to \chi_1 \chi_2$ where the decay $\chi_2 \to \chi_1 e^+ e^-$ is prompt and proceeds through a different mediator which couples to electrons \cite{Izaguirre:2017bqb,Izaguirre:2014dua,Izaguirre:2015zva}.} Here we are agnostic as to the UV completion of such a model, and we are simply aiming for an apples-to-apples comparison between a virtual $S$ or $V$ contributing to $(g-2)_\mu$ and a real $S$ or $V$ emitted from an initial- or final-state muon.
\item \textbf{Phase 2: Thermal muon-philic DM search.} With a larger flux of $10^{13}$ MOT and upgraded detector performance to reject backgrounds at the level of $10^{-13}$, our setup can probe a significant portion of parameter space for which DM is thermally produced through $U(1)_{L_\mu - L_\tau}$ gauge interactions, and $V$ is identified as the gauge boson of this new $U(1)$. 
Such models are inaccessible with both traditional WIMP searches
 \cite{Aguilar-Arevalo:2016ndq,Akerib:2016vxi,Aprile:2017iyp,Agnese:2017njq,Xiao:2017vys,Petricca:2017zdp,Agnes:2018ves} and to most of the
  emerging sub-GeV dark matter search program, which consists of 
 of new direct detection  \cite{Essig:2011nj,Graham:2012su,Essig:2012yx,Hochberg:2015pha,Lee:2015qva,Essig:2015cda,Hochberg:2015fth,Hochberg:2016ntt,Derenzo:2016fse,Essig:2017kqs,Tiffenberg:2017aac,Hochberg:2017wce,Agnes:2018oej,Crisler:2018gci} and fixed target experiments with electron \cite{Izaguirre:2013uxa,Batell:2014mga,Battaglieri:2014qoa,Battaglieri:2016ggd,Izaguirre:2014bca,Mans:2017vej}
 and proton beams \cite{Batell:2009di,Batell:2014yra,deNiverville:2011it,Dharmapalan:2012xp,Aguilar-Arevalo:2017mqx,Izaguirre:2017bqb,Coloma:2015pih,Frugiuele:2017zvx,
 Dobrescu:2014ita}; for a review and summary, see \cite{Battaglieri:2017aum}.


 \end{itemize}

We emphasize that M$^3$ Phase 1 can be completed with minimal modifications to the Fermilab muon source and with only a few months of data-taking. A null result would decisively exclude any new-physics explanation of the $(g-2)_\mu$ anomaly from invisibly-decaying muon-philic particles below 100 MeV. Phase 2 is comparable to the CERN SPS proposal, and in this paper we focus specifically on the advantages of pairing such an experiment with the lower-energy Fermilab muon beam, highlighting the relevance of this search to the thermal DM parameter space. Furthermore, both phases could be implemented as muon-beam reconfigurations of the proposed LDMX experiment with few additional modifications. 

This paper is organized as follows. In section \ref{sec:motivation} we review the physics motivation for our benchmark models; in section \ref{sec:signal} we discuss the characteristics of signal production; in section \ref{sec:expsetup} we describe the 
basic experimental setup and relevant background processes; in section \ref{sec:beamDet} we describe the necessary detector and beam properties; in section \ref{sec:results} we describe the projected sensitivities of our Phase 1 and Phase 2 proposals; finally, in section \ref{sec:conclusion} we offer some concluding remarks. 




\section{Physics Motivation}
\label{sec:motivation}
In this section we present the physics motivation for invisibly decaying muon-specific scalars $S$ or vectors $V$. We begin by reviewing the contributions of vector and scalar particles to $(g-2)_\mu$, and then present a concrete benchmark model with a muon-philic gauge interaction which can be coupled to dark matter. Although this is not the only model that preferentially couples a new force carrier to muons, it serves as a representative example without loss of essential generality. Basic variations away from this example (i.e. the scalar force model in \cite{Chen:2017awl}) feature the same basic degrees of freedom and their 
 signal characteristics are similar to what we consider below. When discussing a generic muon-specific vector mediator, we will use the notation $V$, reserving $Z'$ for the gauge boson coupling the muon to dark matter in our representative model.
 
 \subsection{Simplified Models for $(g-2)_\mu$}
\label{sec:g-2}
The current discrepancy in the anomalous magnetic moment 
is characterized by $a_{\mu}  \equiv \frac{1}{2}(g-2)_\mu$, the
observed value of which differs from the SM theoretical prediction by an amount 
\cite{Patrignani:2016xqp}:
\be
\Delta a_\mu \equiv a_\mu({\rm obs})  - a_\mu({\rm SM}) = (28.8 \pm 8.0) \times 10^{-10}.
\ee
It is well known that for a light scalar $S$ or vector $V$ coupling to the muon,
\be
 g_S S \overline{\mu} \mu \qquad {\rm (scalar)} ~~,~~~
 g_V V_\alpha \overline{\mu} \gamma^\alpha \mu \qquad {\rm (vector)}
\ee
the leading-order contribution to the muon anomalous magnetic moment from scalars is
 \cite{Pospelov:2008zw}
\be
\Delta a^{S}_{\mu}  = \frac{g_S^2}{16\pi^2} \int_0^1 dz \frac{m^2_\mu (1-z) (1-z^2)}{ m_\mu^2 (1-z)^2 +  m_S^2 \, z} \simeq  6.0 \times 10^{-10} \, \left( \frac{g_S}{10^{-4}} \right)^2  ~ ~~ (m_{S} \ll m_\mu),
\ee
and the corresponding expression for vector particles is 
\be
\Delta a^{V}_{\mu}  = \frac{g_V^2 }{4\pi^2} \int_0^1 dz \frac{ m^2_\mu z (1-z)^2}{ m_\mu^2 (1-z)^2 +  m_V^2 \, z} \simeq  1.6 \times 10^{-9} \, \left( \frac{g_V}{10^{-4}} \right)^2  ~ ~~ (m_{V} \ll m_\mu).
\ee
Note that pseudoscalar and axial-vector couplings contribute to $a_\mu$ with opposite sign, pushing the theoretical value farther away from the measured value. As a result we only consider parity-even mediators, scalars $S$ or vectors $V$. 

For the purposes of our Phase 1 search, we make no attempt to describe a complete theoretical model of $S$ or $V$, and take the observed discrepancy in $a_\mu$ as positive evidence for a new particle which can be probed in beam dump experiments. For Phase 2, we can define a well-motivated region of parameter space and make a connection to thermal dark matter in a particular representative model, which we describe below.
  
  \subsection{A Complete Vector Model: $U(1)_{L_\mu-L_\tau}$}
For our Phase 2 study, we extend the SM to include the anomaly-free $U(1)_{L_\mu-L_\tau}$ gauge group under which  
 which $\mu, \tau$ and their corresponding neutrino flavors couple to a new gauge boson $\zp$. 
 The Lagrangian for this scenario is 
\be \label{eq:lag1}
{\cal L} = {\cal L}_{\rm SM}  - \frac{1}{4} { F^{\prime} }^{\alpha \beta}  F^{\prime}_{\alpha \beta}    + \frac{m_{\zp}^2}{2} {Z^{\prime} }^\alpha  Z^{\prime}_\alpha  - Z^\prime_\alpha J^{\alpha}_{\mu-\tau}   , ~~
\ee
where $  F^{\prime}_{\alpha \beta}  \equiv \partial_\alpha {Z^\prime}_\beta -  \partial_\beta {Z^\prime}_\alpha$  is the field strength tensor and
 $m_{\zp}$ is the gauge boson mass; we assume that the gauge symmetry is spontaneously broken in the IR, but that states responsible for that breaking
are sufficiently decoupled that their effects are negligible at the GeV scale. 
The $\mu -\tau$ current in Eq.~(\ref{eq:lag1}) is 
\be
\label{eq:Jmutau}
J^{\alpha}_{\mu - \tau} = g_{\mu - \tau} \left(    \bar \mu\gamma^\alpha \mu +  \bar \nu_\mu \gamma^\alpha P_L \nu_{\mu}  -
							  \bar \tau\gamma^\alpha \tau -  \bar \nu_\tau \gamma^\alpha P_L \nu_{\tau}    \right), 
\ee
where $g_{\mu - \tau}$ is the gauge coupling and $P_L \equiv \frac{1}{2}(1-\gamma^5)$ is the left projection operator.
The rest frame partial widths for $\zpr$ decays are
\be
\Gamma(\zp \to f f) =  \frac{\alpha_{\mu - \tau} m_{\zp}}{3} \left( 1 + \frac{2m^2_f}{m^2_{\zp}} \right)     \sqrt{1 - \frac{4 m_f^2}{m^2_{\zp} }}  ~~~~~       (f = \mu, \tau)~~ ,~~
\Gamma(\zp \to \nu_i\nu_i) = \frac{\alpha_{\mu - \tau} m_{\zp}}{6} ,~~~
\ee
where $\alpha_{\mu - \tau} \equiv g_{\mu - \tau}^2/\ 4\pi$. In the absence of additional decay channels,  for $m_{\zp} < 2 m_\mu$ the $\zp$ 
will always decay invisibly to $\mu$ and $\tau$  neutrinos and for $m_{\zp} > 2 m_\mu$ the invisible branching ratio is 
approximately $1/3$.  If the $\zp$ can also decay to lighter hidden sector states (see Sec.~\ref{sec:add-DM}), it is possible to have 
a large branching fraction to invisible states throughout the parameter space if the hidden sector particles couple more strongly to
the mediator.

Note this model modifies neutrino interactions, so there are strong limits from neutrino trident scattering
via the process $\nu N \to \nu N \mu^+\mu^-$ which exclude this interpretation of the 
$(g-2)_\mu$ anomaly for $m_{\zp} \gtrsim 400$ MeV.  Furthermore, for $m_{\zp} \lesssim$ few MeV, there are 
strong bounds from cosmology since a $\zp$ in this mass range decays to neutrinos after the latter have decoupled
from the photons of the SM thermal bath. Such a late decay heats the neutrinos relative to the photons, thereby
predicting $\Delta N_{\rm eff} \sim 3$. Thus, the current Planck limit of
 $N_{\rm eff} = 3.15 \pm 0.23$ \cite{Ade:2015xua} rules out this mass range unless the $\zp$ is too feebly coupled
 to ever thermalize with SM particles in the early universe. However, avoiding thermalization requires $g_{\mu - \tau} \ll 10^{-5}$, which
 falls far short of resolving the $(g-2)_\mu$ anomaly, so our primary
  focus for the remainder of this work will be on the range $ \MeV \lesssim m_{\zp} \lesssim 400 \, \MeV$.

 \subsection{Including Dark Matter }
 \label{sec:add-DM}

In addition to the model in Eq.~(\ref{eq:lag1}), we can also include dark matter  charged  under $U(1)_{L_\mu-L_\tau}$, which 
adds the additional interaction  ${\cal L} \supset  - Z^\prime_\alpha J^{\alpha}_{\chi} $ to the Lagrangian in
Eq.~(\ref{eq:lag1}). For a variety of benchmark DM candidates $\chi$ with mass $m_\chi$, the current 
coupled to the $\zp$ is 
\be
J_\chi^\mu= g_\chi \times  
\begin{cases}
                        i \chi^* \partial_\mu  \chi + h.c. ~ & {\rm Complex ~ Scalar} \\
                             \overline \chi_1 \gamma^\mu \chi_2 + h.c.  ~  & {\rm Pseudo\!\!-\!\!Dirac~ Fermion} \\
              \frac{1}{2}  \overline \chi \gamma^\mu \gamma^5 \chi ~  &{\rm Majorana~ Fermion} \\
                     \overline \chi \gamma^\mu \chi ~ & {\rm Dirac~ Fermion} \\
                    \end{cases}
\ee
where $g_\chi = g_{\mu - \tau} q_\chi$ is the $\zp$ coupling to DM and $q_\chi$ accounts for the possibility that DM does not carry unit charge under $U(1)_{L_\mu - L_\tau}$. 
As long as the DM is vector-like under the gauge extension, $q_\chi$ is a free parameter, so we 
adopt the convention from Eq.~(\ref{eq:Jmutau}) which assigns unit charge to $\mu, \tau$, and their neutrinos. Generically, loops of $\mu$ and $\tau$ will generate kinetic mixing between the $Z'$ and the photon (this scenario and its implications for electron beam experiments are explored further in Refs.~\cite{Gninenko:2018tlp,Bauer:2018onh}), but this can always be cancelled by another UV contribution to the kinetic mixing parameter. In this work we assume that \emph{no} kinetic mixing is present in the IR, and the $Z'$ does not couple to charged particles other than muons and taus.

\begin{figure}[t!] 
\hspace{-0.35cm}
\includegraphics[width=7.7cm]{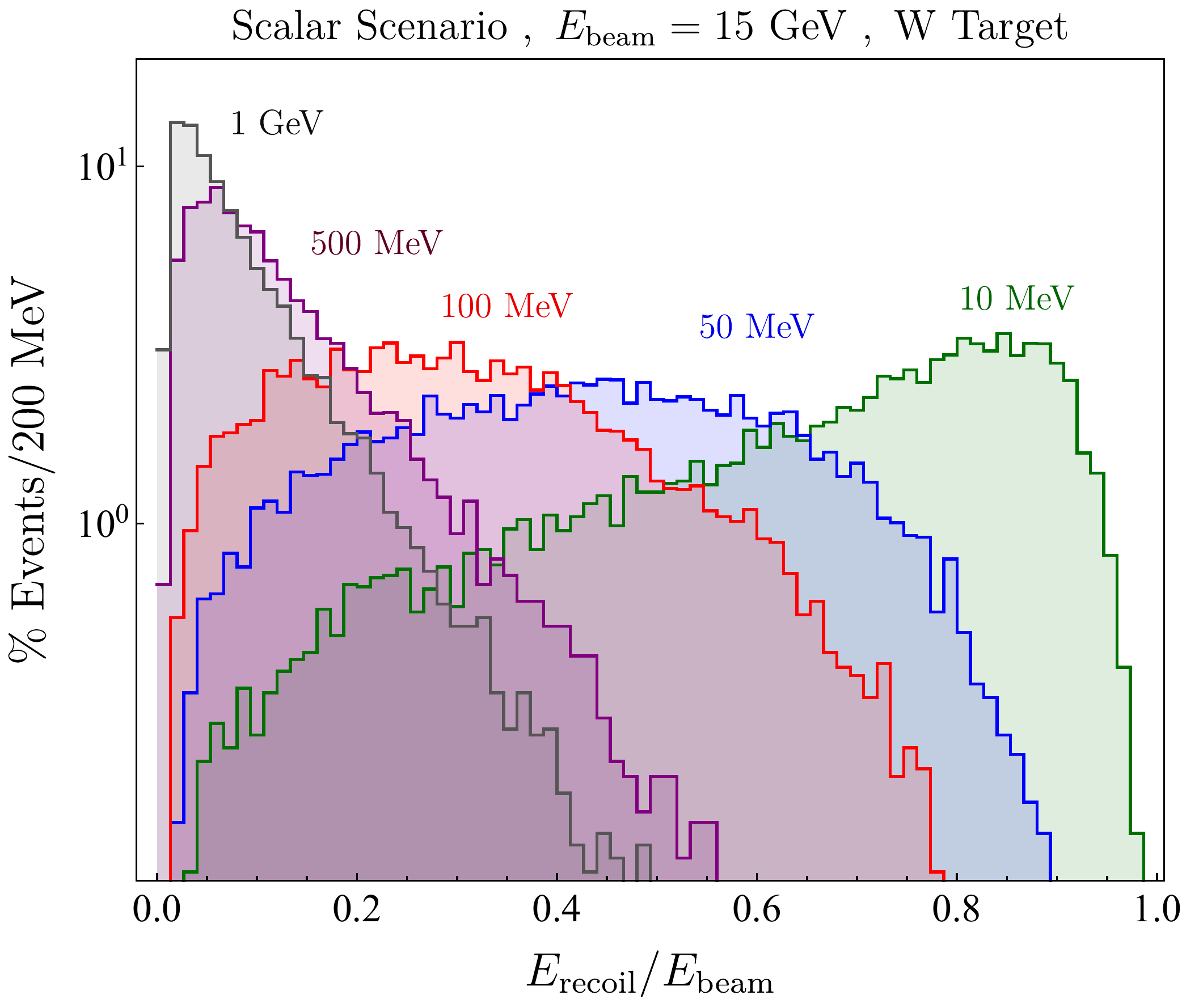}~~
\includegraphics[width=7.7cm]{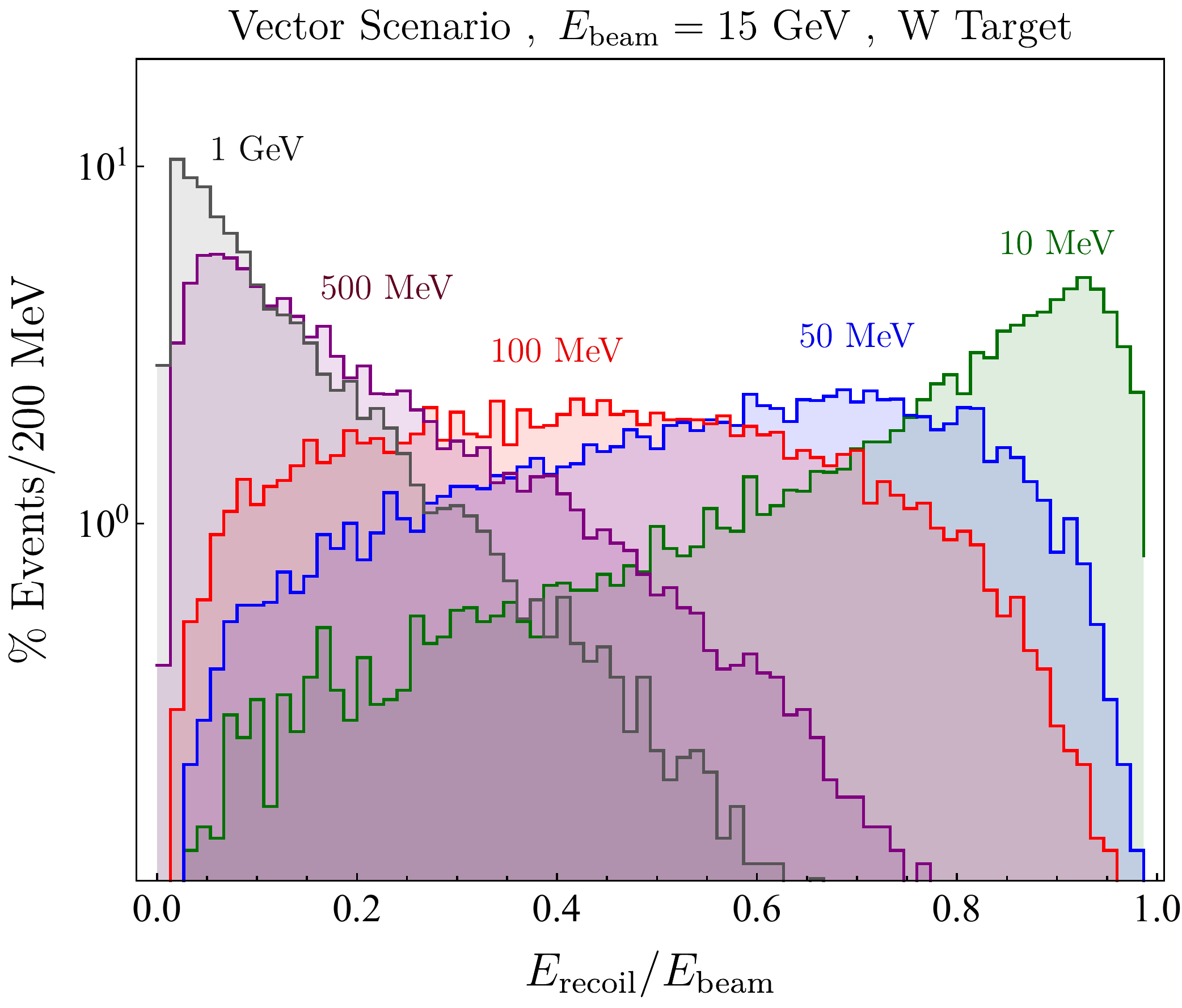}
   \caption{Energy distributions for signal events for simplified phenomenological models with invisibly decaying scalar $S$ (\emph{left}) and vector $V$ (\emph{right}) particles that
   couple only to muons. Each histogram corresponds to a different choice of scalar or vector mass.}
\vspace{0cm}
\label{fig:SignalDistEnergy}
\end{figure}

\begin{figure}[t!] 
\hspace{-0.35cm}
\includegraphics[width=7.7cm]{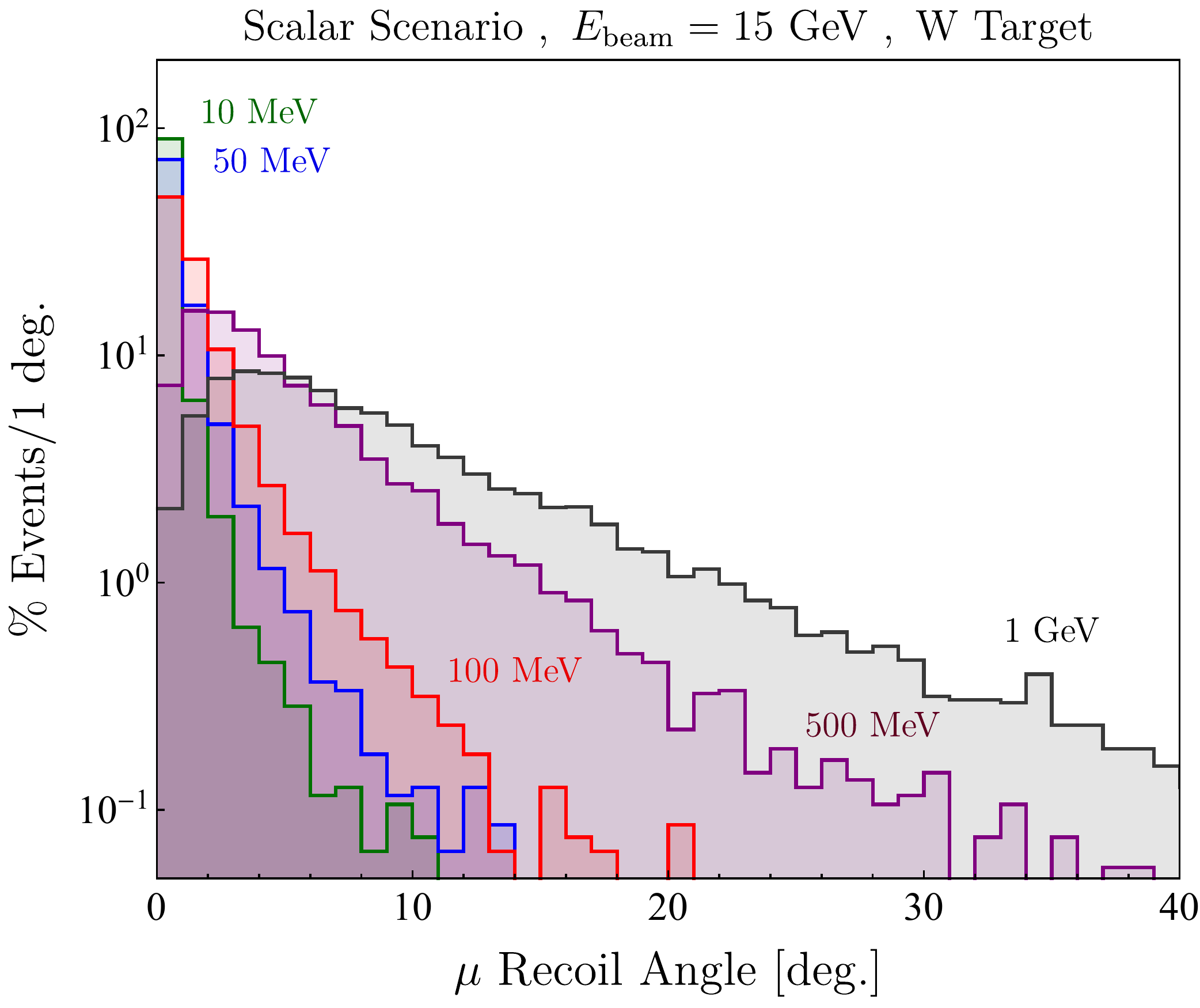}~~
\includegraphics[width=7.7cm]{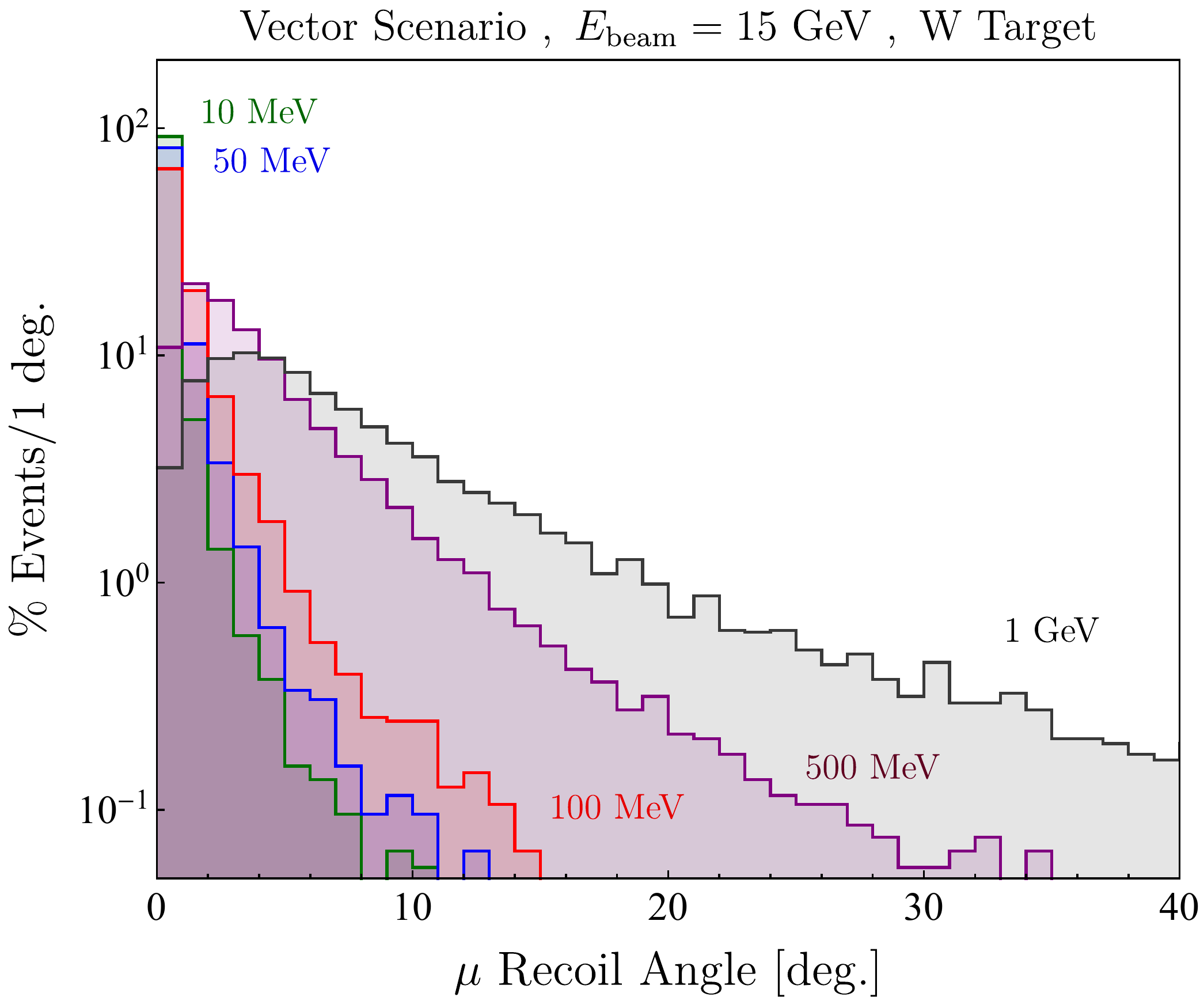}
   \caption{Angular distributions for signal events for phenomenological simplified models with invisibly decaying scalar $S$ (\emph{left}) and vector $V$ (\emph{right}) 
   particles coupled to muons.  }
\vspace{0cm}
\label{fig:SignalDistAngle}
\end{figure}

In the early universe, the $L_\mu - L_\tau$ gauge boson $Z'$  mediates interactions that maintain thermal equilibrium between dark and visible matter. In the absence of a 
particle asymmetry in the dark sector,  the relic abundance is determined by the dark-visible interaction rate at the freeze-out temperature  $T_{\rm fo} \sim m_\chi/20$. 
For $m_\chi > m_{\zp}$ the abundance is set primarily by $\chi \chi \to \zp \zp$ {\it secluded} annihilation which depends only on $g_\chi$ and is independent
of the $\zp$ coupling to SM particles. In the opposite regime $m_\chi < m_{\zp}$, the abundance is determined by direct annihilation, $\chi\chi \to {\zp}^* \to \mu^+ \mu^-, \tau^+\tau^-, \nu_i \nu_i$, 
the cross section for which is proportional to the combination ($g_\chi g_{\mu - \tau})^2$. In this case, even for $g_\chi \sim 1$ 
 near the perturbative unitarity bound, there is still a minimum $g_{\mu - \tau}$ required to yield the observed abundance at late times. 

To define relic targets for the experimental program, we consider in the $m_{\zp }  \gg m_\chi$ for a Dirac $\chi$ particle. Defining
the dimensionless variable $y \equiv g_\chi^2 g_{\mu-\tau}^2 (m_\chi/m_{\zp})^4$, the cross section and abundance 
have the approximate scaling  
 \be
\langle  \sigma v \rangle \simeq \frac{3 g_\chi^2 g^2 m_\chi^2}{\pi^2 m_\zp^2} = \frac{3y}{\pi m_\chi^2}  ~ \implies ~
 \Omega_\chi h^2 \sim 0.1 \left( \frac{3 \times 10^{-9}}{ y}\right)   \biggl( \frac{m_\chi}{ \rm GeV}\biggr)^2 ,   
 \ee
 where the factor of $3$ accounts for annihilation channels and the requisite $y$ from this estimate 
 agrees well with the full numerical results presented in Fig \ref{fig:thermal-fig}. As we will show in Sec.~\ref{sec:results}, there is a region of parameter space where $U(1)_{L_\mu - L_\tau}$ can explain both thermal DM and the $(g-2)_\mu$ anomaly, but the parameter space for thermal DM is a priori unconstrained by $(g-2)_\mu$.




\section{Signal Kinematics}
\label{sec:signal}

In this section we discuss the physics of the new mediator radiated from a muon beam, noting some important features of the signal production cross section and kinematics. In the limit where the beam energy is much greater than both $m_{S,V}$ and $m_\mu$, it is appropriate to use the generalized Weizsacker-Williams approximation for the cross section for mediator production, where we also keep track of the mass of the beam particle \cite{Liu:2016mqv,Liu:2017htz,Chen:2017awl}. For scalar $S$ production we have
\be
 \left. \frac{d \sigma}{dx} \right |_S  \simeq \frac{g_S^2 \alpha^2 }{4\pi} \chi_S \beta_S \beta_\mu \frac{x^3 \left [ m_\mu^2(3x^2 - 4x + 4) + 2m_S^2(1-x) \right]}{\left [ m_S^2(1-x) + m_\mu^2 x^2 \right]^2} , \label{eq:WWscalar}
 \ee
 and a corresponding expression for $V$ production
 \be
  \left. \frac{d \sigma}{dx} \right |_V  \simeq \frac{g_V^2 \alpha^2 }{4\pi} \chi_V \beta_V \beta_\mu \frac{2x \left [ x^2m_\mu^2(3x^2 - 4x + 4) - 2m_V^2(x^3 - 4x^2 + 6x - 3) \right]}{\left [ m_V^2(1-x) + m_\mu^2 x^2 \right]^2},
\label{eq:WWvector}
\ee
where $x = E_{S,V}/E_{\rm beam}$, $\beta_{S,V} = \sqrt{1 - m_{S,V}^2/(x E_{\rm beam})^2}$, $\beta_\mu = \sqrt{1 - m_\mu^2/E_{\rm beam}^2}$, and
\be
\chi_{S,V} = \int_{t_{\rm min}}^{t_{\rm max}} \, dt \, \frac{t - t_{\rm min}}{t^2} G_2(t) \simeq \int_{m_{S,V}^4/(4 E_{\rm beam}^2)}^{m_{S,V}^2 + m_\mu^2} \frac{t - m_{S,V}^4/(4E_{\rm beam}^2)}{t^2} G_2(t),
\ee
with $G_2$ the target form factor.

\begin{figure}[t!] 
\hspace{-0.25cm}
\includegraphics[width=7.4cm]{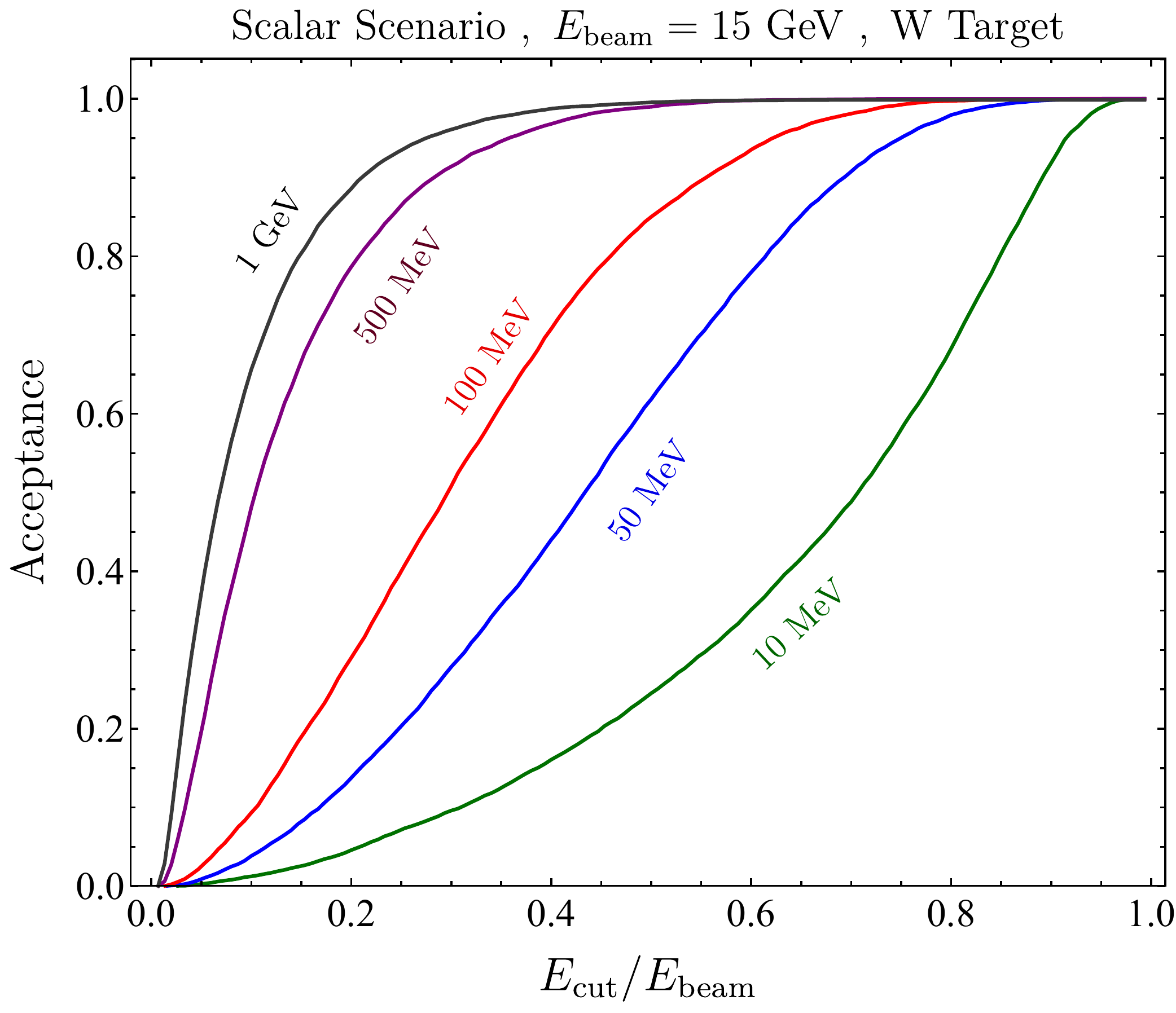}~~~~
\includegraphics[width=7.4cm]{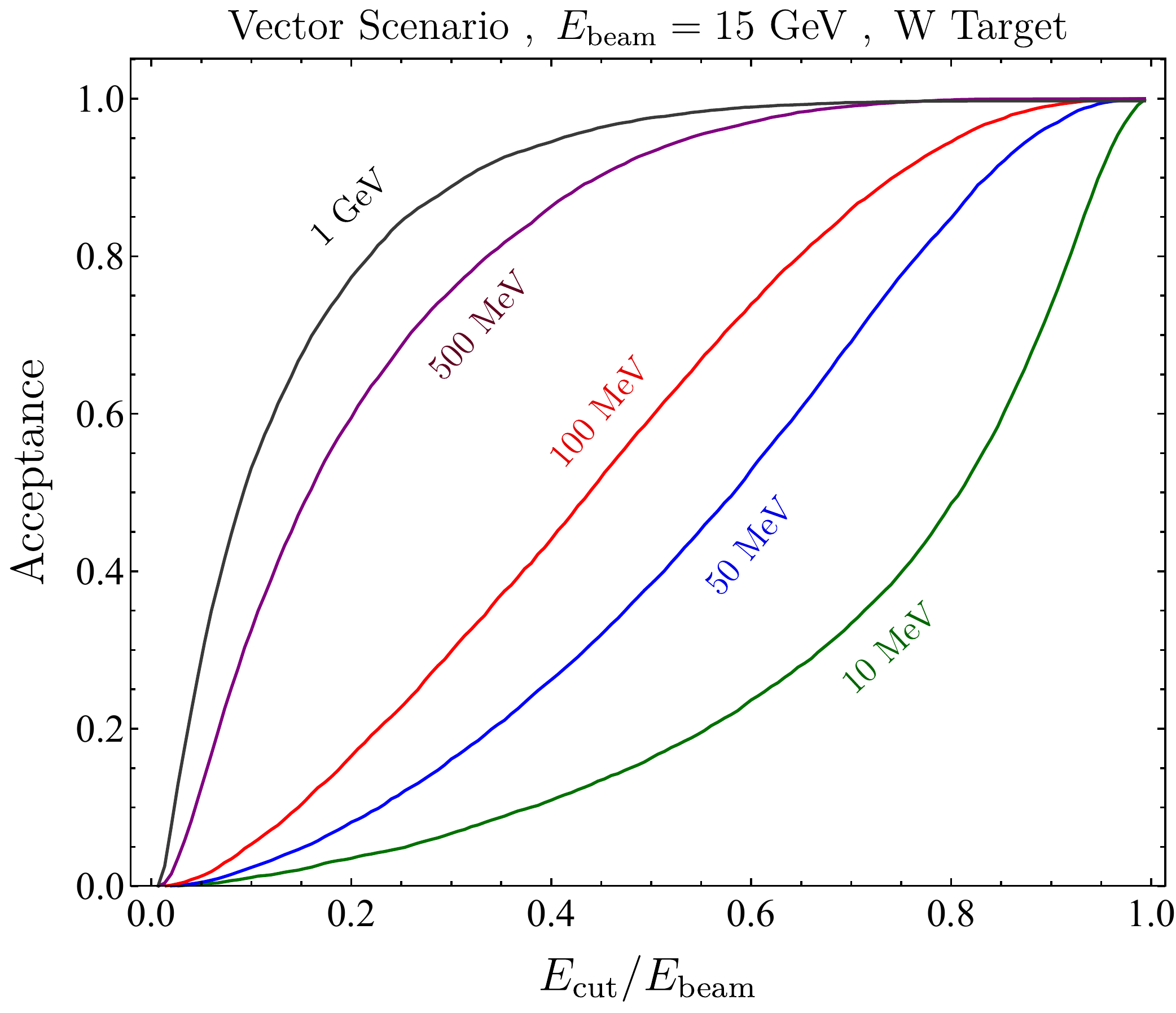}
\caption{\label{fig:accept} 
Signal acceptance for radiated scalar $S$ (\emph{left}) and vector $V$ (\emph{right}) particles as a function of the fractional cut on the maximum muon recoil energy for various choices
of $S/V$ masses.  
}
\vspace{0cm}
\end{figure}

\begin{figure}[t!] 
\hspace{-0.5cm}
\begin{center}
\includegraphics[width=7.7cm]{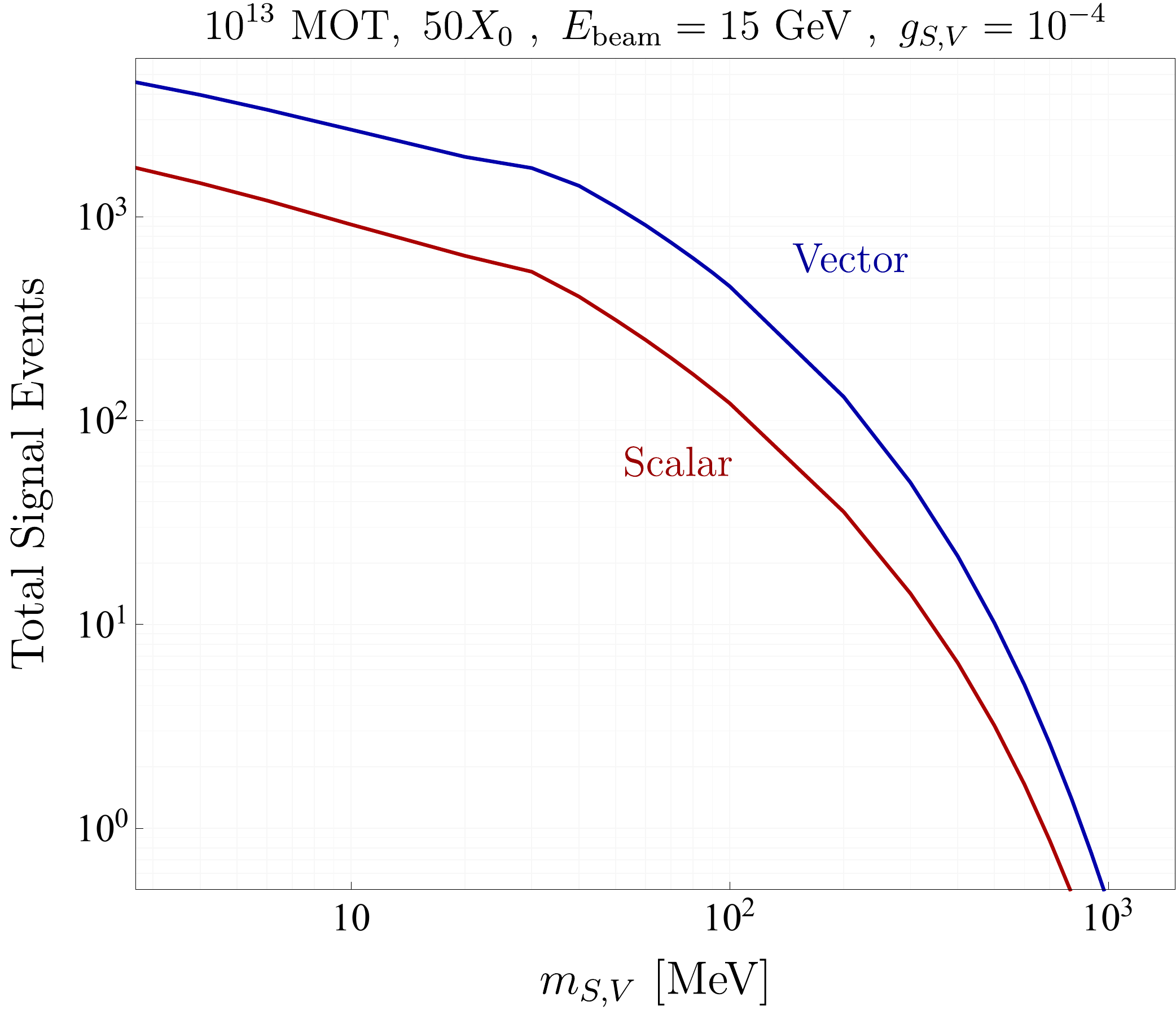}~~~
\end{center}
\caption{\label{fig:cross} 
Signal event yield for radiative $S$ or $V$ production in muon-tungsten coherent scattering with a 15 GeV incident beam energy, $10^{13}$ muons on target, and a target of thickness $50~X_0$. Here we have normalized to a reference coupling $g_{V,S} = 10^{-4}$ which is roughly the value required to explain the $(g-2)_\mu$ anomaly; see Sec.~\ref{sec:g-2} and Fig.~\ref{fig:g-2fig}.}
\vspace{0cm}
\end{figure}

As has been noted in previous studies, and as shown here in Fig.~\ref{fig:SignalDistEnergy}, the energy spectrum differs markedly depending on whether $m_{S,V} < m_\mu$ (in which case $S/V$ is soft and the spectrum closely resembles ordinary QED bremsstrahlung) or $m_{S,V} > m_\mu$ (in which case $S/V$ takes most of the beam energy, as is familiar from electron beam studies). The same is true to a lesser extent for the angular spectrum (Fig.~\ref{fig:SignalDistAngle}), where the outgoing muon is peaked in the forward direction for light $S/V$ but can be emitted at wider angles for heaver $S/V$. However, we also note here that both the shape of the energy spectrum and the inclusive cross section are largely independent of beam energy for small $m_{S,V}$. This can be seen directly from Eqs~(\ref{eq:WWscalar}--\ref{eq:WWvector}), where $d\sigma/dx$ is only a function of the recoil energy fraction $x$ and not $E_{\rm beam}$ in the limit $m_{S,V} \to 0$ and $m_\mu \ll E_{\rm beam}$; we have also confirmed this with \texttt{Madgraph} simulations.\footnote{We thank Chien-Yi Chen and Yiming Zhong for providing us with the modified \texttt{Madgraph} files which implement fixed-target scattering on tungsten targets.} Therefore, the reach of a muon beam experiment at low masses will be determined almost entirely by the cut on recoil energy carried by the outgoing muon, required to keep backgrounds to a tolerable level. In Section~\ref{sec:beamDet} we will find that $E_{\rm cut} = 9 \ \GeV$ with $E_{\rm beam} = 15 \ \GeV$ allows us to achieve zero background, while still corresponding to $\mathcal{O}(1)$ acceptance over much of the desired mass range. Fig.~\ref{fig:accept} shows the signal acceptance as a function of recoil cut, and Fig.~\ref{fig:cross} shows the inclusive cross section for a beam energy of 15 GeV, for both $S$ and $V$. The vector cross section is roughly a factor of 3 larger for the same value of the coupling, due almost entirely to the spin sum over the 3 polarization states for a massive vector.




\section{Experimental Setup}
\label{sec:expsetup}

In this section, we introduce the basic experimental concept which takes advantage of the missing momentum technique~\cite{Bjorken:2009mm}.  
We describe the general layout of the detector setup, leaving a dedicated discussion of the detector technology for Section~\ref{sec:beamDet}. After introducing the experimental technique, we discuss the main background processes.

\subsection{Missing momentum technique}

The missing momentum technique probes the dark bremsstrahlung production of new, invisibly decaying particles produced in $\mu N \to \mu N \displaystyle{\not}{E}$ processes, as illustrated in Fig.~\ref{fig:feyn}, from an incident muon beam on a fixed nuclear target.
The particle of interest ($S$ or $V$) is radiated off of the muon as initial- or final-state radiation via interaction with the nucleus and, by assumption, promptly decays to dark matter or neutrinos, thereby yielding missing energy and momentum.  The technique requires identifying and measuring the momentum of individual muons; the experimental signature is an outgoing muon emerging from the target with significant momentum loss compared to the incident muon, with no additional visible energy deposited in the target material or in the downstream ECAL/HCAL,  as illustrated in Fig.~\ref{fig:experiment}.

The signal production cross section depends on the incoming beam energy and the number of radiation lengths, $X_0$, of the target. Good muon momentum resolution for both the incoming and outgoing tracking system is crucial. A high-field dipole magnet, along with its fringe field, surrounds the tracking region in order to provide a good momentum measurement. The target should be active to detect background processes from hard muon scattering in the target. Downstream of the target, we have additional electromagnetic and hadronic calorimeters to capture muon energy loss from standard QED bremsstrahlung and 
possible photon-nucleon and muon-nucleon hard processes which produce hadrons. The angular spectrum of the signal process, shown in Fig.~\ref{fig:SignalDistAngle}, informs the geometry of the tracker and calorimeters.

Two key features separate this experimental proposal from previous studies~\cite{Gninenko:2014pea,Chen:2017awl}: the energy of the incident muons (15 GeV as opposed to 150 GeV in previous studies), and the technology, based on the proposed LDMX experiment, which is well-suited to this energy scale and does not rely on measurement of visible decay products of $S$ or $V$. This design allows for a more compact experimental setup while still achieving the necessary detector performance. In this paper we consider an incident 15~GeV muon beam as a typical beam energy for the experiment, but a range of comparable muon energies is reasonable and our setup does not rely crucially on the precise value of the beam energy.

\subsection{Backgrounds}
\label{sec:bkgs}

The relevant background processes for electron beams were enumerated in \cite{Izaguirre:2014bca}, and for muon beams in \cite{Gninenko:2014pea}. Here, we summarize the various categories of backgrounds and how their effects are modified in the case of a muon beam. The main difference between muon and electron beams is that muons are minimum ionizing particles (MIPs), and will continuously lose energy passing through a thick target. Energy loss of the muon without detectable signals in the target from electronic interactions, typically referred to as $dE/dx$, for a 15~GeV muon through a radiation length $50~X_0$ of tungsten absorber material is roughly 550~MeV.  Large fluctuations of the energy loss can cause a signal-like event topology. However, these events are often accompanied by other particles produced in a hard process, and can be vetoed. A study of this effect and discussion of vetoes in our setup is given in Section~\ref{sec:det}.

\subsubsection{Beam-related backgrounds}

\begin{itemize}

\item {\bf Muon energy spread:} The Fermilab muon source derives from decays in flight of a monoenergetic 32 GeV pion beam. Prior to delivery to the the target, the beam passes through an iron shield where most of the surviving pions are absorbed. Muons lose energy and scatter in the shield, resulting in a spread of muon momenta exiting the shield, from 10 to 30 GeV with a significant low-energy tail below 3 GeV from hard bremsstrahlung in the shield~\cite{JensenPriv}. In our proposal, the tagging tracker is designed to measure the momentum of each incoming muon. While it will be difficult to trigger on this momentum and neglect any events with low-energy muons, such events can be discarded in the analysis phase. The tagging tracker must be designed such that the momentum measurement is sufficiently accurate to identify muons with energy below the desired recoil threshold to an accuracy of $10^{-10}$ for Phase 1 and $10^{-13}$ for Phase 2; we discuss the performance of the tracker in Sec.~\ref{sec:det}.

\item {\bf Pion contamination:} Pions which decay in the target region can fake a signal because the pion mass is so close to the muon mass, and the muon beam is not mono-energetic. Specifically, the initial beam particle momentum is measured solely by track curvature in the tagging tracker region, and pions of slightly lower energy are degenerate with muons of slightly higher energy. Pions which decay in the bulk of the tagging tracker region can be distinguished from muons by the kink in the track curvature resulting from the lower-energy muon decay product. The background events will be pions which decay between the second-to-last tagging tracking layer and the first recoil tracking layer. Since $X_0 = 0.35 \ {\rm cm}$ for tungsten, the minimum thickness of the target region is $\ell = 50~X_0 = $ 17.5 cm. Assuming the pions all have energy 32 GeV, $\gamma c \tau = 1790 \ {\rm m},$ and so the probability of decay in the target region is $P_{\rm decay} \approx \ell/\gamma c \tau = 10^{-4}$. Thus the pion contamination must be less than $10^{-6}$ for Phase 1, and less than $10^{-9}$ for Phase 2. The current design of the test beam facility includes 12 feet of iron, which should be sufficient to provide a beam purity of $10^{-6}$ for Phase 1 (see Sec.~\ref{sec:Phase1}). For Phase 2, an additional hadron absorber can be placed between the beam source and the tagging tracker, as suggested in \cite{Gninenko:2014pea}.

\end{itemize}

\subsubsection{Reducible backgrounds}

\begin{itemize}
\item {\bf Single bremsstrahlung:} $\mu N \to \mu N \gamma$. This background results from a hard photon escaping detection in the calorimeter. For a sufficiently thick calorimeter, the single-photon bremsstrahlung $e N \to e N \gamma$ was shown to be negligible for LDMX, and will be further suppressed by $(m_e/m_\mu)^2$ for muon beams.

\item {\bf Bremsstrahlung-initiated hadronic events:} $\mu N \to \mu N \gamma$, $\gamma N \to$ hadrons. Many hadronic events will deposit a significant amount of energy into the active target or will produce easily identifiable secondaries. 
Rare processes such as $\gamma p \to \pi^+ n$, $\gamma N \to (\rho, \omega, \phi)N \to \pi^+ \pi^- N$, and $\gamma n \to n \bar{n} n$ can fake signal events depending on the probability of the hadronic calorimeter missing the final-state hadrons. Again, since these processes are photon-initiated, they are suppressed by $(m_e/m_\mu)^2$ compared to the case of electron beams, and thus can be mitigated with relatively mild background rejection; with veto inefficiencies $\lesssim 0.01$ for charged pions and $ \lesssim 0.1$ for neutrons, these processes will give less than 1 fake event per $10^{15}$ incident muons. This informs the detector performance requirements for Phase 2.

\item  {\bf Muon pair production:} $\mu N \to \mu N \mu^+ \mu^-$ or $\mu N \to \mu N \gamma$, $\gamma N \to N \mu^+ \mu^-$. This rate is small for Phase 1 luminosity, but is the dominant reducible background for the Phase 2 muon beam. Indeed, the kinematics of the Bethe-Heitler ``trident'' diagram are peaked in the region of phase space where one muon is produced collinear with the incoming muon \cite{Essig:2010xa}, but this background is reducible since the majority of the time the remaining muon can be tagged, and/or the presence of two MIPs can be seen in the tracker. We discuss the performance of the tracker further in Sec.~\ref{sec:LDMXmumod}, and include a detailed discussion of the pair production background in Appendix~\ref{app:Trident}.
\end{itemize}

\subsubsection{Irreducible backgrounds}

\begin{itemize}

\item {\bf Irreducible neutrino pair production:} $\mu N \to \mu N \nu \overline{\nu}$. The only single process resulting in real missing energy relevant for muon beams is $Z$-mediated neutrino pair production, but the small cross section at 15 GeV beam energies renders this rate negligible even for Phase 2.

\item {\bf Moller + CCQE:} $\mu e \to \mu e$, $ep \to n \nu_e$. For this process, the incident muon scatters a target electron, which acquires a large fraction of the incident beam
energy and then undergoes a charged-current quasi-elastic (CCQE) scatter off a proton. The resulting final state contains one low energy muon and a majority of the beam
energy carried away by the $n$ and $\nu$. If the neutron is not vetoed, this process is an irreducible background, whose probability per muon is conservatively
estimated to be 
\be
P_{ \mu e \to \mu e \rm + CCQE}=   ( \alpha\, n_e  G_F    \ell)^2  \frac{m_p}{2 m_e}  \log \frac{ E_{ \mu} }{ E_{\rm cut}} \simeq 5 \times 10^{-15},
\ee
where, for a tungsten target, the electron density is $n_e = 4.6 \times 10^{24} \, \cm^{-3}$, the radiation length is $X_0 = 17.5$ cm, 
and $E_{\rm cut} =  E_{\rm \mu}/2$ is signal trigger threshold -- see Appendix~\ref{app:CCQE} for a derivation and discussion. Thus, this background is irrelevant for the luminosities we consider in this paper.  

\end{itemize}

Finally, we mention a class of background processes which are irreducible for electron beams but negligible for muon beams. CCQE with final-state neutrinos, $e N \to N' \nu_e$, accompanied by a $\gamma$ or $\pi^0$ in the final state, can pose a significant background when the $\gamma$ or $\pi^0$ fakes an electron. However, the analogous process for muon beams, $\mu N \to N' \nu_\mu + (\gamma, \pi^0)$, would be easily vetoed by the target and ECAL signals.

\subsubsection{Target background rates for Phases 1 and 2}

\begin{figure}[t!] 
\vspace{-0.2cm}
\center
\includegraphics[width=10cm]{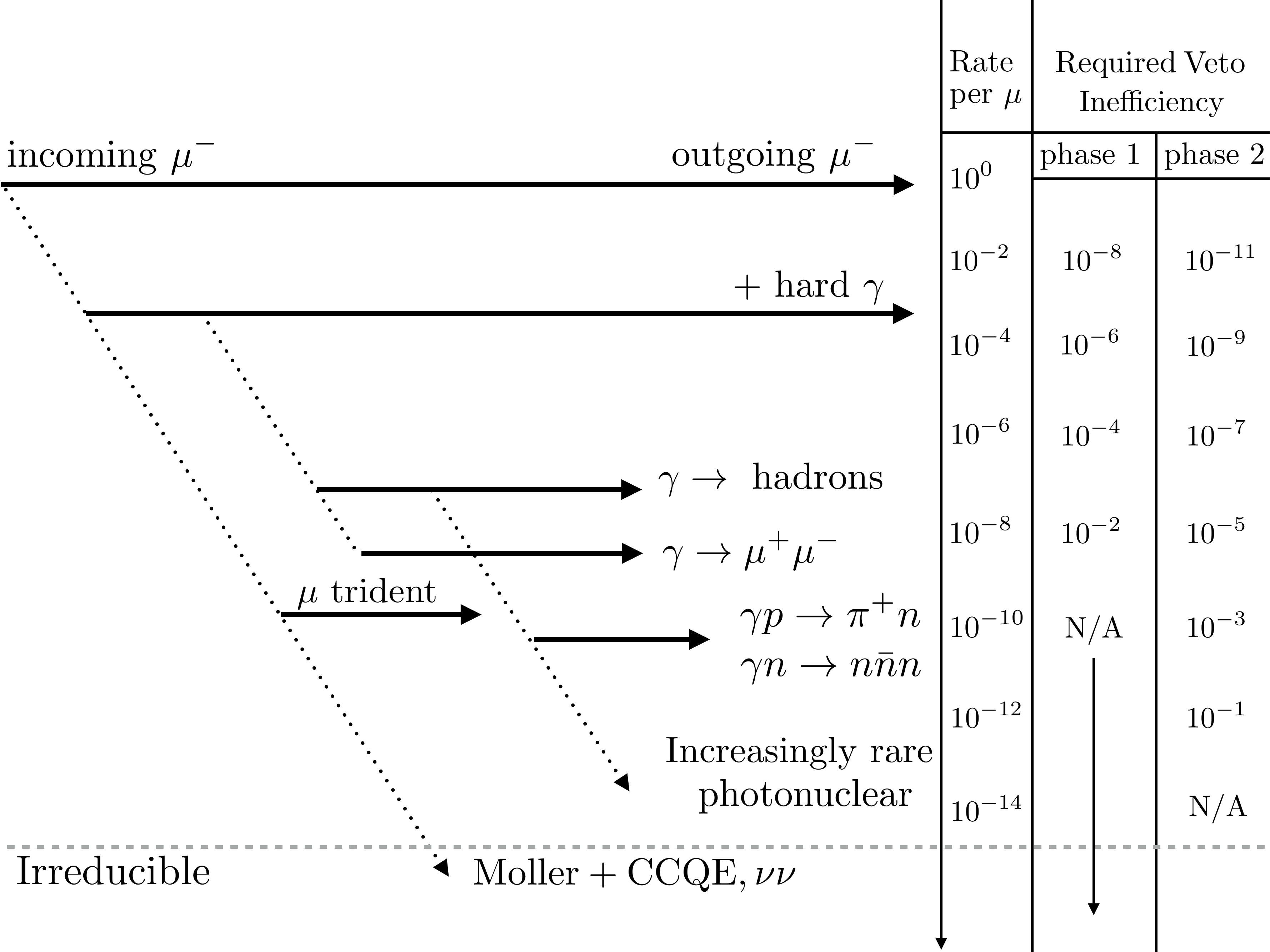} 
\caption{\label{fig:backgrounds} 
Relative rates of various background processes and the required veto inefficiencies for a zero background
experiment for phases 1 and 2.}
\vspace{0cm}
\end{figure}

The dominant background processes and relative rates are shown in Fig.~\ref{fig:backgrounds}, along with the maximum veto inefficiencies for Phase 1 and Phase 2 required to achieve the required sensitivity. 
Because of the reduced bremsstrahlung rate for muon beams and overall smaller total integrated muons on target, there are milder requirements on the hadronic vetoes than for electron beams. 
In summary, we expect that a muon beam experiment based on the LDMX detector concept and technology can be made background-free with $10^{13}$ muons. 
While detailed studies of the experimental backgrounds are beyond the scope of this work, we have performed a \texttt{GEANT} \cite{Agostinelli:2002hh} simulation of muons on an example target geometry to verify the rates of the rarer reactions, such as photonuclear and muon-nuclear interactions.  
Discussions of these studies are given in Sec.~\ref{sec:LDMXmumod}.

\section{Detector and Beam Parameters}
\label{sec:beamDet}

The experimental setup, described in the previous section, requires the ability to identify and track individual muons.
Thus a low-current, high-repetition rate muon beam would be ideal, with per-bucket occupancies of $\mathcal{O}$(few) muons.  
The sensitivity of the experiment to $g_{S,V}$ is proportional to $\sqrt{N_\mu}$, so acquiring as many muons as possible on target is paramount. To maintain an experimental setup similar to LDMX, which is relatively compact and therefore low-cost, the beam energy should be $\mathcal{O}$(several) GeV. This drives the recoil muon $p_T$ resolution.

Given these general beam requirements we discuss a specific realistic example in order to define benchmark scenarios for the physics reach of such an experiment.  We take as our example the possibility of a muon beam at the Fermilab Accelerator complex~\cite{Brown:2013idd}.

\subsection{An LDMX-like Detector}
\label{sec:det}

We now discuss the most important differences between an LDMX-like detector for a muon beam experiment and its original design with an electron beam. More details of the LDMX design can be found in \cite{Mans:2017vej}.

\subsubsection{LDMX design}
The LDMX detector is very similar to the schematic laid out in Fig.~\ref{fig:experiment} except that the target area is much thinner, approximately 0.1 $X_0$.  
The tagging tracker upstream of the target measures the incoming electron momentum.  The recoil tracker, along with the electromagnetic calorimeter (ECAL), measures the outgoing electron momentum.  The tagging and recoil trackers employ a silicon microstrip detector with strip readout pitch of 60~$\mu$m.  
The tagging tracker and target are inside the bore of a 1~T dipole magnet while the recoil tracker is in the fringe field of the magnet.
The electromagnetic calorimeter is a high granularity tungsten-silicon sample calorimeter with a hexagonal geometry and hexagon cell size of 1~cm.
The silicon-based calorimeter is required primarily for high electron energy resolution and radiation tolerance as the electromagnetic calorimeter is effectively an instrumented beam dump.  The hadronic calorimeter is a steel-scintillator sampling calorimeter with a high sampling fraction for high neutral hadron detection efficiency. 
The detector is designed to handle incoming electron rates of 40-200~MHz and requires readout electronics accordingly.

The tracking system is based on technology in use at the HPS experiment~\cite{Celentano:2014wya,Adrian:2015hst} and the electromagnetic calorimeter is based on technology being developed for the CMS ECAL endcap HL-LHC upgrade~\cite{CMSCollaboration:2015zni}.

\subsubsection{M$^3$ modifications}
\label{sec:LDMXmumod}

The primary change to an LDMX-like detector paired with a muon beam is the target area, which now must be much thicker.  
Let us nominally consider a 50~$X_0$ thick target region.
The most important design feature for such a thick target is that it is active in order to detect muon energy loss from SM interactions within the target. 
We propose, as a simple extension of the detector technology, to use the same modules for the target in M$^3$ that are employed in the ECAL in LDMX. 
The high granularity and high sampling fraction make it a good candidate technology for high-efficiency detection of hard bremsstrahlung and photonuclear interactions.  Other technologies such as LYSO scintillator are viable options, but here we consider the high granularity silicon-based calorimeter as the primary option as it is a minimal change to the detector technologies needed for the experiment.
Compactness is important for fitting inside the magnet, both in the transverse and longitudinal directions.  One particular challenge is accomodating the cooling, power, and electronics infrastructure for the active target. 

In $\rm{M}^3$, the ECAL no longer provides information on the muon momentum, so the muon momentum is measured exclusively by the recoil tracker. As a result, the recoil track momentum resolution is paramount to the experiment. A simple estimate for the momentum resolution in the recoil tracker gives
\begin{equation}
\frac{\sigma_p}{p} \approx \frac{\sqrt{8/n} \sigma_x}{h} \frac{p}{qLB_y},
\end{equation}
where $p$ is the muon momentum, $n$ is the number of measurements by the tracker (nominally 6), $\sigma_x$ is the hit resolution ($60/\sqrt{12}~\mu$m), $LB_y$ is the length in the (fringe) magnetic field ($\sim$0.1~T-m) and $h$ is the lever arm.  
For a lever arm of 0.25~m, this gives approximately 0.8\% momentum resolution.
This resolution drives the cut on the muon recoil momentum. The primary signal selection is defined by this cut combined with a minimal amount of energy loss in the active target from the muon (see below). The recoil track momentum resolution can be improved fairly simply by a factor of 2--4 by increasing the recoil track lever arm (at a loss of angular coverage) and/or with a more powerful magnet (2~T).

\begin{figure}[t!] 
\center
\includegraphics[width=0.48\linewidth]{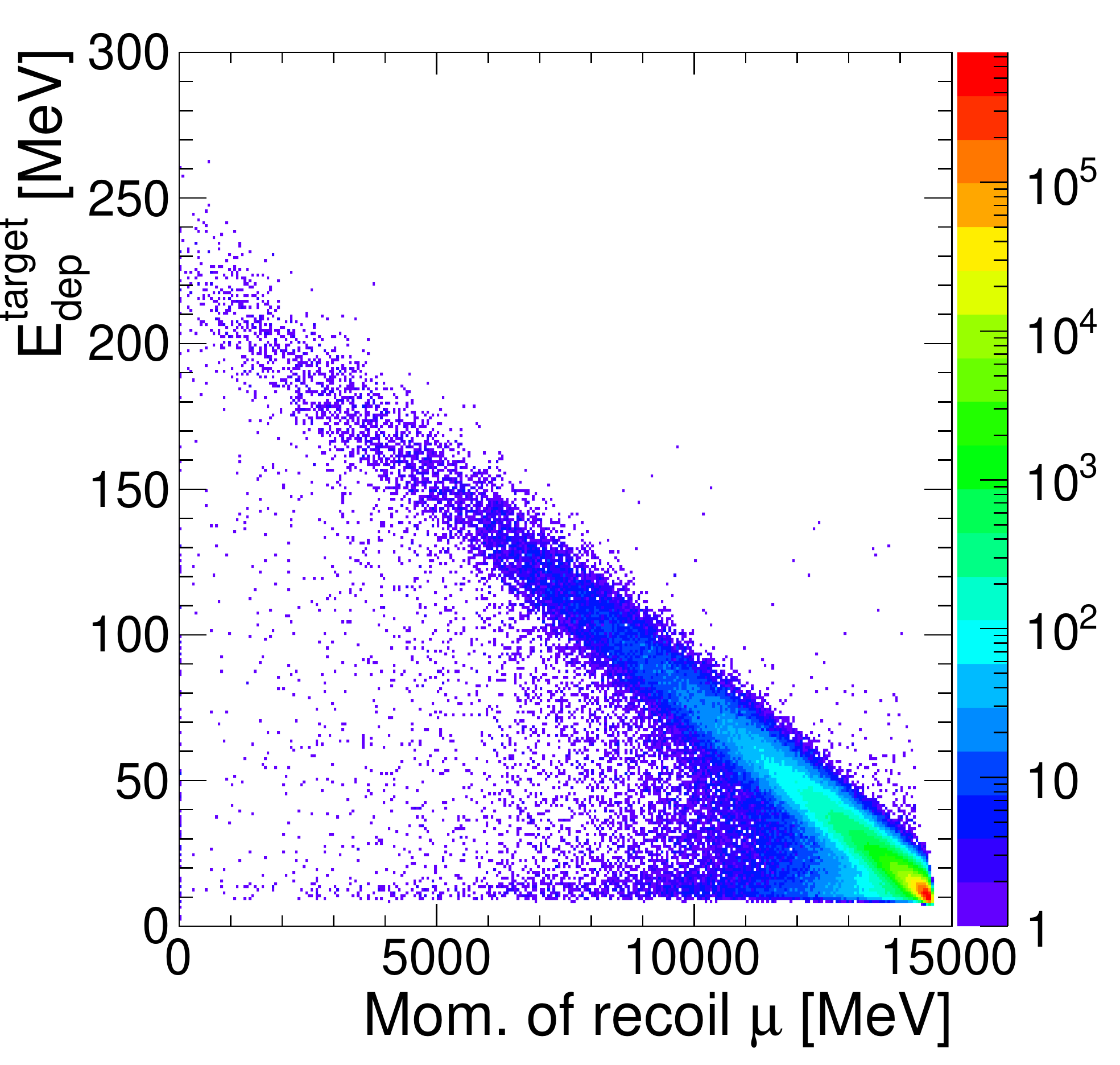}
\includegraphics[width=0.48\linewidth]{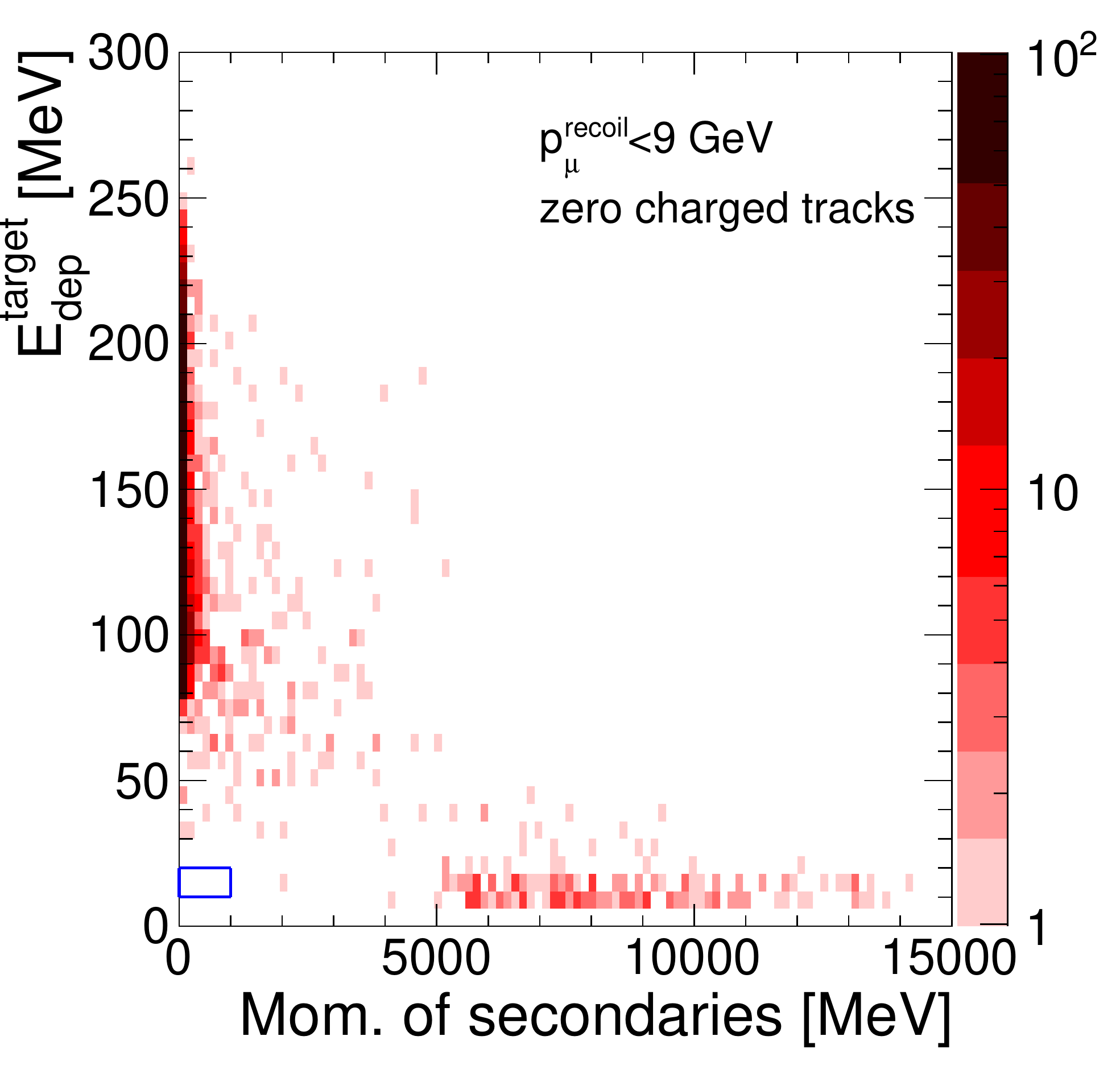}
\caption{\label{fig:dedx} 
(\emph{Left}) Distribution of recoil momentum of muons leaving the target versus energy deposited 
in active elements of the target. (\emph{Right}) Distribution of energy deposited in the active elements
of the target versus non-muon (secondary) momentum leaving the target for events with recoil muon momentum less than $9$~GeV and 
zero charge particle leaving the target at an angle less than $40^\circ$. The blue box demarcates the simple signal selection described in the text. 
}
\end{figure}

In order to find a reasonable selection for the signal, we study the energy loss of a muon traversing a 50~$X_0$ active target modeled after the LDMX electromagnetic calorimeter. 
The modeling is performed using the LDMX software suite~\cite{ldmxsw} based on  the \texttt{GEANT} simulation framework~\cite{Agostinelli:2002hh}. 
We model a sample of $10^{7}$ muons with 15~GeV of incident energy on a tungsten-silicon sampling calorimeter where tungsten absorber layers are 1~$X_0$ per layer.  
We then examine the energy of the outgoing muon compared to the amount of energy deposited in the silicon.  
We can cut on the amount of energy deposited in the silicon to eliminate events where the muon underwent a hard interaction in the calorimeter, causing significant muon energy loss.  
The results are shown in Fig.~\ref{fig:dedx}.

In  Fig.~\ref{fig:dedx} (left), one can see that the incoming 15~GeV muon typically loses about 500~MeV of energy traversing the target.
The important metric is the fluctuations of energy loss in the target without any significant energy deposited in the silicon in the target. 
The silicon energy deposit is shown on the $y$-axis,
and we see that below 12.5~GeV of recoil momentum, there are relatively few events with less than 20~MeV energy deposited. 
Because all events have a minimum ionizing track traversing the target, very few events populate the region below 10~MeV.  
The signature of signal
events, which consist of a single minimum ionizing muon, will be $10-20$~MeV of energy deposited in the active layers of the target. 
Given the muon momentum 
resolution of $\sim1\%$ , we define a signal region which requires a muon outgoing 
momentum of $ < 9~\rm{GeV}$, 40 standard deviations away from the incoming beam energy.  
We combine this with a simplistic selection on the energy  of $10-20~\rm{MeV}$ deposited in the silicon. 

After defining this signal selection, we illustrate in Fig.~\ref{fig:dedx} (right) the amount of non-muon momentum exiting the back of the target at an angle of less than $40^\circ$.
This is plotted against the energy deposited in the silicon on the $y$-axis.
Because additional charged particles downstream of the target are easily identified, those events are vetoed and what is plotted in Fig.~\ref{fig:dedx} (right) are the remaining events. 
Relatively high non-muon momentum downstream of the target is also easily vetoed and so the signal region is defined by the blue box which denotes only energy from a minimum-ionizing particle deposited in the silicon
and no non-muon momentum  exiting the back of the target.
The plot shows no background events in the signal region for the statistics generated. 

Going beyond this simple signal selection, we could use additional information on MIP track shape and anomalous cell hits 
to further reject muons undergoing significant energy loss and not leaving much energy in the target. 
While we do not consider them in this work, these are important additional handles that can be used to reject background processes.
Beyond the target and the recoil tracking systems, not much modification is needed to the nominal LDMX design. 
The ECAL and HCAL system will continue to be important in rejecting debris from photon interactions occurring late in the target or producing hadrons, though (as described in Sec.~\ref{sec:bkgs}) these backgrounds are suppressed with respect to the electron beam scenario. 
The trigger system may need to be modified slightly for muon rates above $\sim$100~kHz, which is the current LDMX event rate capacity; this is most relevant for Phase 2, and we discuss this situation further below.   
In addition, the current detector configuration should be able to handle multiple muons per event, however, more than $\mathcal{O}\rm{(several)}$ would result in hit confusion from combinatorics and reduced signal efficiency.







\subsection{Fermilab Accelerator Complex}
\label{sec:phases}

We now discuss a potential accompanying beamline for the M$^3$ detector at the Fermilab Accelerator Complex.
The experiment imagines a two-phase setup where, with lower total integrated luminosity (muons on target), the first phase will 
cover the $(g-2)_\mu$ parameter space and the second will probe the thermal DM parameter space.

\subsubsection{Phase 1: MTest beamline}
\label{sec:Phase1}

The primary Main Injector beamline provides a proton beam of 120~GeV with an RF frequency of 53~MHz~\cite{Brown:2013idd}.
The Fermilab Test Beam Facility (FTBF)~\cite{fbtf-page} beam is extracted from the Main Injector beam via resonant extraction over 4.2~s spills.
The beam is then sent to the MTest or MCenter facilities which receive beams of various energies and particle types.

At the energies we require, the MTest beamline can produce a 10--30 GeV muon beam which is produced from pion decays at 32~\GeV.
For a pion beam of 32~\GeV, the beamline has a $>80\%$ composition of charged pions.
In muon mode, pions are converted to muons where the final expected rate of $10^5$ muons per spill~\cite{Denisov:2016djw}.
Pion contamination of the MTest beamline is controlled by an upstream hadronic absorber.
For the M$^3$ experiment, charged pion contamination should be at the level of $10^{-6}$ which is achievable with the available space in the beamline at MTest.
The transverse beam size is roughly a few centimeters in $x$ and $y$.
The time between spills is approximately one minute.
Over one week of continuous running, this sums to approximately $1 \times 10^9$ muons on target (MOT).  Such a performance is achievable with little modification to the Fermilab Accelerator Complex. Therefore we consider the MTest beamline an excellent candidate for Phase 1 of the experiment which probes a large portion of the allowed region for $(g-2)_\mu$.
In Phase 1, we plan for the following experimental parameters: 
\begin{itemize}
\item muon energy = 15~GeV 
\item muon intensity = $10^{10}$ MOT (muons on target)
\item Target Thickness = $50~X_0$ (about 25 cm for tungsten/silicon)
\end{itemize} 

This program could be achieved on a timescale of roughly a few months.
There are no additional detector requirements beyond what is described in Section~\ref{sec:det}.
Given the estimated background rates in Fig. \ref{fig:backgrounds} and the muon repetition rate at MTest, a zero-background experiment is achievable.

\subsubsection{Phase 2: Neutrino (NM4) Beamline}

In order to achieve higher incident pion rates, improved radiation shielding for the MTest target area is required. A candidate would be a similar beamline, called ``Neutrino Line'', which similarly performs resonant extraction of beam from the Main Injector but sends the protons down another beamline which currently hosts the SeaQuest experiment.
This beamline is specified to handle $10^6 - 10^7$ muons per spill which could greatly increase the sensitivity of the M$^3$ technique.
However, the current Neutrino Line would require modifications to beamline itself to enable muon production. 
With this higher integrated luminosity of muons on target, we can now achieve enough MOT to probe the thermal DM phase space.  

In Phase 2, we plan for the following experimental parameters:  
\begin{itemize}
\item muon energy = 15~GeV 
\item muon intensity = $10^{13}$ MOT
\item Target Thickness = $50~X_0$
\end{itemize} 

This program would be achieved in a longer timescale than Phase 1 depending on the achievable muon instantaneous luminosity.
With a higher muon instantaneous luminosity, a trigger is needed to reduce the event rate to a more manageable 100~kHz.
This will require additional R\&D into the hardware for fast track reconstruction in order to keep the readout rate manageable. 
However, even in this scenario, background rates are expected to be manageable.

\section{Projected Sensitivities}
\label{sec:results}

\begin{figure}[t!] 
\hspace{-0.5cm}
\includegraphics[width=7.6cm]{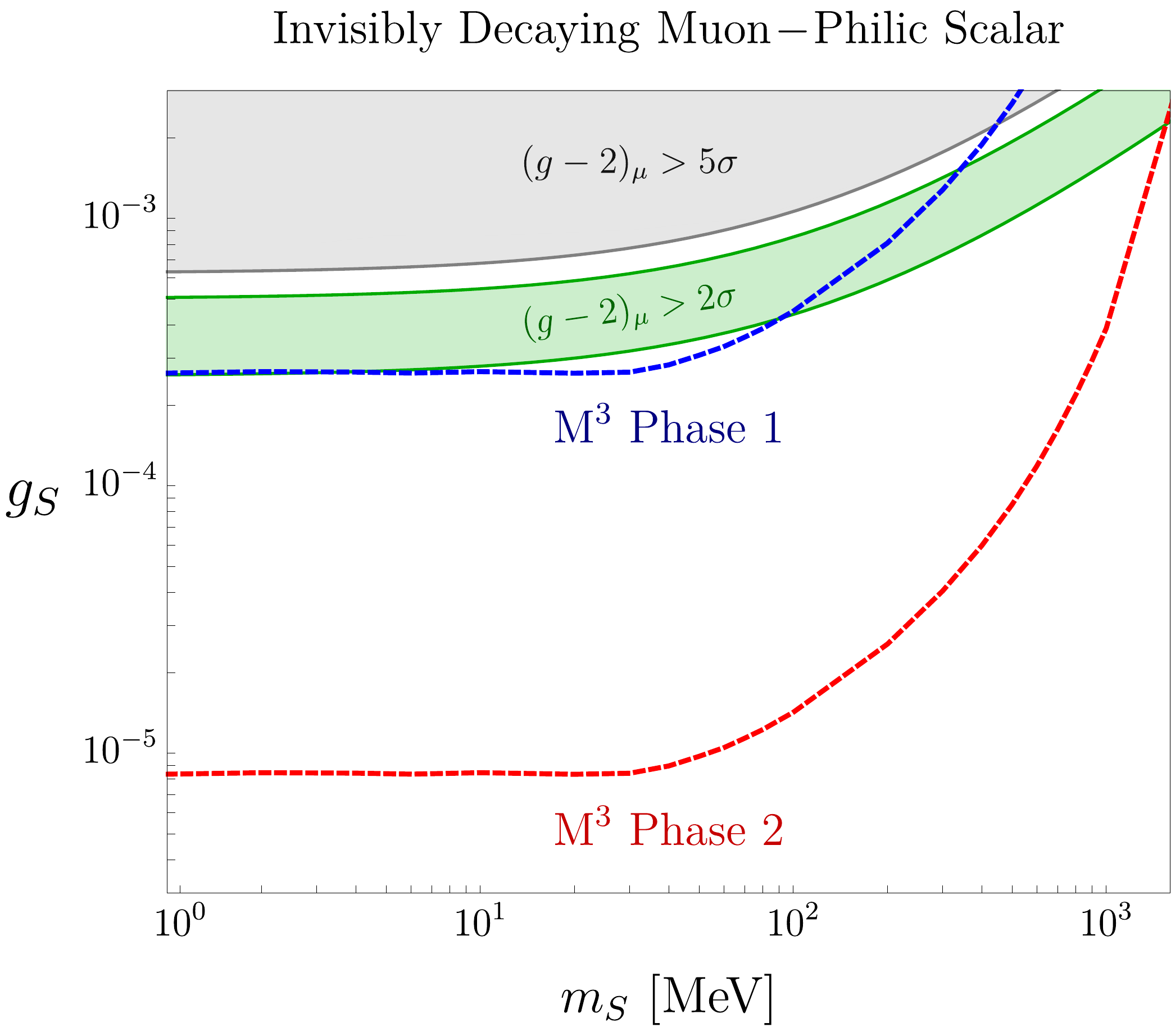}~~~
\includegraphics[width=7.6cm]{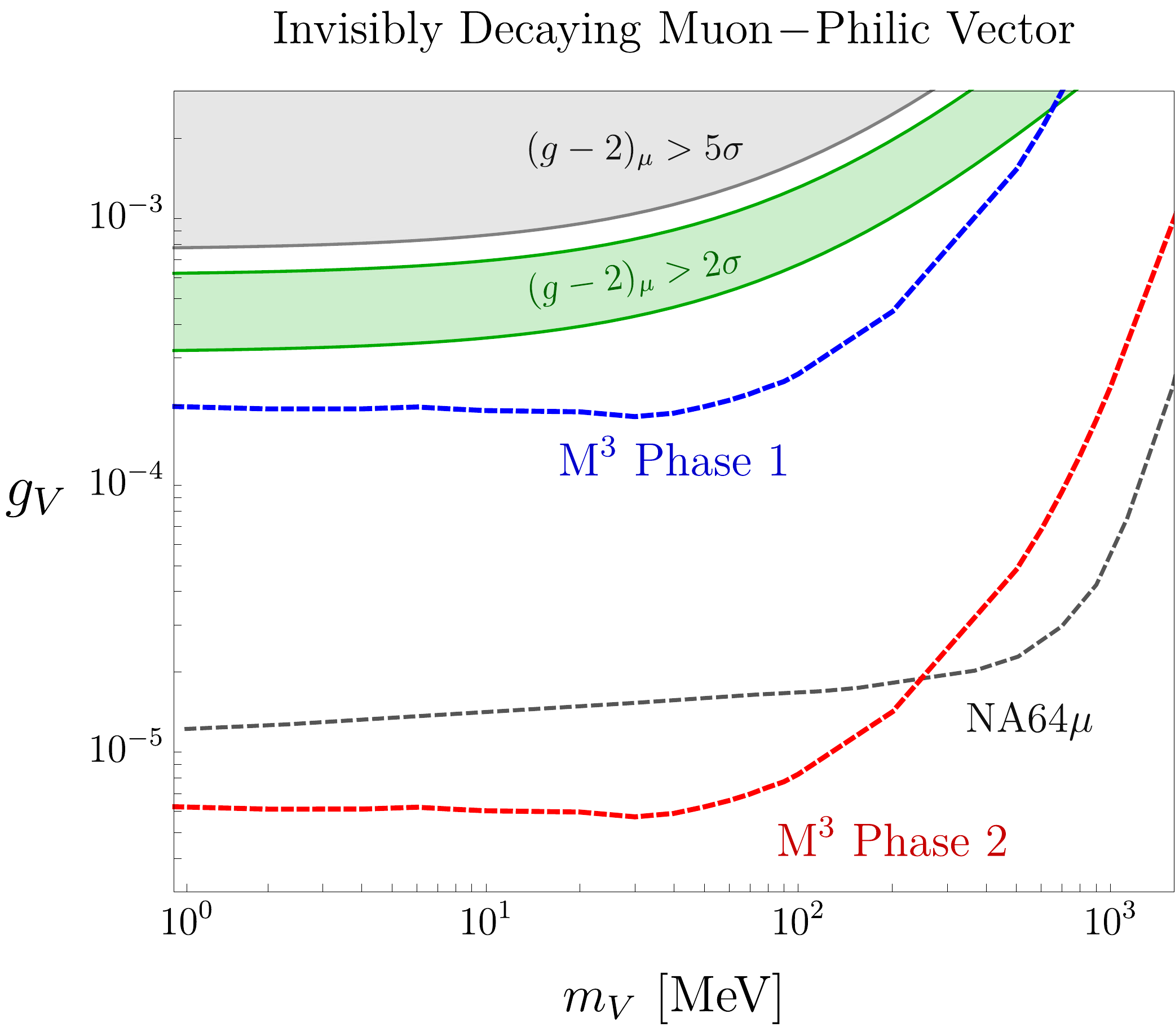}~~~
   \caption{Parameter space for a muon-philic scalar $S$ (\emph{left}) or vector $V$ (\emph{right}) particle as described in Sec \ref{sec:g-2}. The green
   bands represent the parameter space for which such particles can reconcile the $(g-2)_\mu$ anomaly to within $2\sigma$ of the measured value. Also
   shown are the M$^3$ projections for Phases 1 and 2 involving the Fermilab test beam facility and the Neutrino (NM4) beamline respectively (see Sec.~\ref{sec:phases} for more details). For the vector plot, the gray dashed curve is the NA64$\mu$ projection for an invisibly decaying vector particle, which uses the projections in \cite{Gninenko:2014pea}, but rescales the region
  $m_V > 2m_\mu$  to ensure BR($V \to~\rm{invisible}) = 1$; a comparable analysis for scalars at NA64$\mu$ would also cover new parameter space. Note also that for $m_{V,S} <$ 1 MeV, the new invisibly decaying particles will be 
  in thermal equilibrium with the Standard Model during BBN and increase $N_{\rm eff}$, so we do not consider this regime.
  }
\vspace{0cm}
\label{fig:g-2fig}
\end{figure}

In this section, we discuss the sensitivity of the experimental proposal in the zero background regime. 
For our signal projection, we assume a 50 $X_0$ tungsten target and 
generate a sample of $\mu N \to \mu N (S,V)$ simulated events for various choices of $V,S$ masses using a modified version of \texttt{MadGraph} which
 includes nuclear target form factors (see Sec \ref{sec:signal}). We also impose energy and angular cuts
 to keep only those events in which the outgoing muon passes through the forward tracking layers ($p_z > 0$ to avoid backwards
 scattering) and satisfies $E < E_{\rm cut}  = 9$ GeV. For most values of $m_{S,V}$ that we consider, these cuts yield an order-one signal acceptance
as shown in Fig.~\ref{fig:accept}.

\subsection{Phase 1: $(g-2)_\mu$ sensitivity}

Fig.~\ref{fig:g-2fig} shows the target parameter space for light new physics contributions to $(g-2)_\mu$. The grey region is excluded because the contribution to $(g-2)_\mu$ is greater than $5\sigma$ from the measured value, while green region would resolve the $(g-2)_\mu$ anomaly to within $\pm2\sigma$. We see that Phase 1 can completely exclude all new-physics explanations for $(g-2)_\mu$ at mass scales below 100 MeV, and higher for vector mediators and/or looser cuts which could still allow for a background-free search. For comparison, we also show the reach for Phase 2, which can extend the exclusion region to 1 GeV for both scalars and vectors.

\subsection{Phase 2: $U(1)_{L_\mu - L_\tau}$ thermal DM sensitivity}

\begin{figure}[t!] 
\hspace{-0.7cm}
\includegraphics[width=8.cm]{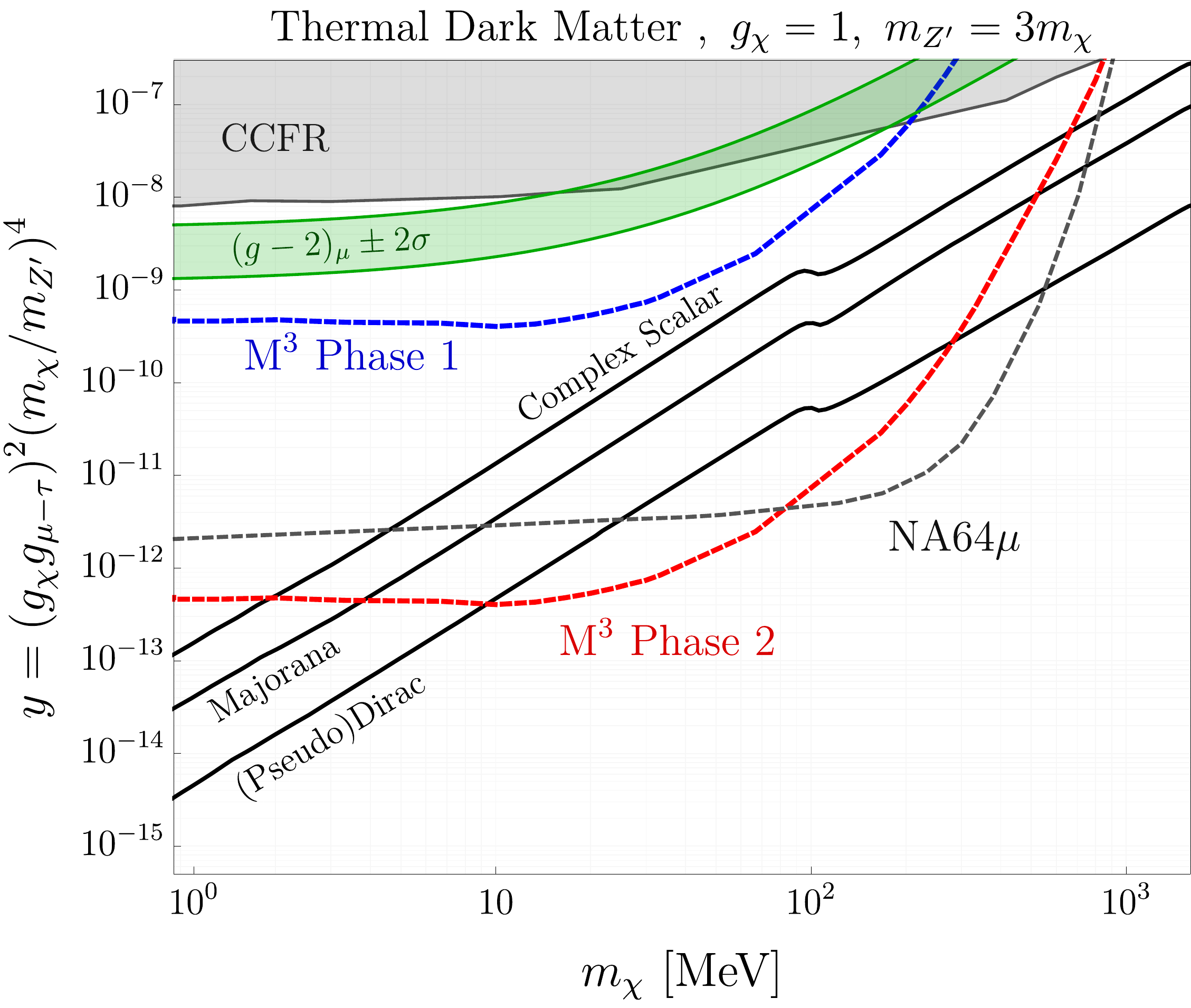}~~~
\includegraphics[width=8.cm]{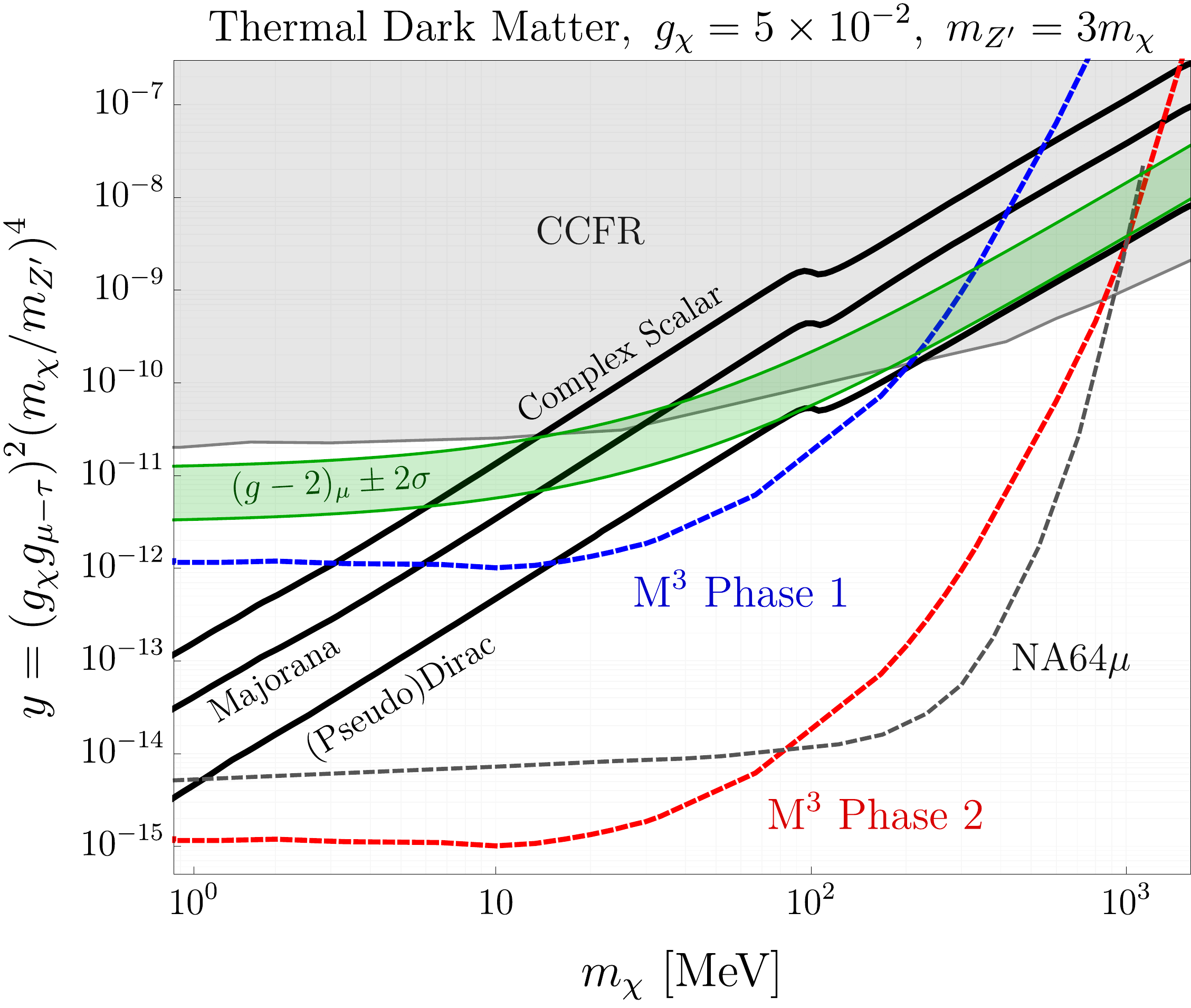} 
   \caption{ Parameter space for predictive thermal DM charged under $U(1)_{L_\mu - L_\tau}$, for DM charges near the perturbativity limit (\emph{left}) or smaller such that the $(g-2)_\mu$ region overlaps with the thermal relic curves (\emph{right}). Here the relic
  abundance arises through direct annihilation to SM particles via $s$-channel $\zp$ exchange.The vertical axis is the product of couplings that sets the relic abundance for a given choice of DM mass and spin (see Appendix~\ref{app:relic-density}). Also plotted are constraints from the neutrino trident process from the CCFR experiment \cite{Mishra:1991bv,Altmannshofer:2014pba} and projected limits from NA64 \cite{Gninenko:2016kpg}. Note that there are also bounds on $m_\chi = \mathcal{O}(\MeV)$ from $\Delta N_{\rm eff.}$ that arise from $\chi\bar\chi \to \nu \nu$ annihilation during BBN; these bounds differ depending on the choice of DM candidate spin \cite{Boehm:2013jpa,Nollett:2014lwa} and are not shown here. For the pure Dirac scenario, the annihilation process $\chi\bar\chi \to \mu^+\mu^-$ is $s$-wave, so this process is ruled out by CMB energy injection bounds for $m_\chi > m_\mu$ \cite{Ade:2015xua}.   }
  \label{fig:thermal-fig}
\vspace{0cm}
\end{figure}

Fig.~\ref{fig:thermal-fig} shows the target parameter space for thermal relic DM with a $L_\mu - L_\tau$ mediator. The vertical axis plots the dimensionless variable $y = g_\chi^2 g_{\mu-\tau}^2 (m_\chi/m_{Z'})^4$ which controls the DM annihilation rate, and the black curves represent the unique value of $y$ for each $m_\chi$ which results in the correct DM relic abundance (see appendix \ref{sec:appendix-density}), for DM a complex scalar, Majorana fermion, or (pseudo)-Dirac fermion (see Sec.~\ref{sec:add-DM}). The left panel shows the scenario $g_\chi = 1$ near the perturbativity limit, which corresponds to the weakest possible bounds on this model, while the right panel shows the case $g_\chi = 5 \times 10^{-2}$. In the latter case, there is a region of parameter space compatible with both thermal dark matter and $(g-2)_\mu$, which can be probed by Phase 1, with the entire viable parameter space for thermal DM probed by Phase 2.\footnote{See also \cite{Calibbi:2018aa} for other models relating thermal DM to $(g-2)_\mu$.}  Even for the pessimistic case $g_\chi = 1$, a large portion of the parameter space is accessible to Phase 2. We emphasize that muon beam experiments like M$^3$ are the only terrestrial experiments which can probe such a muon-philic model of DM; direct detection signals are absent, and high-energy collider production cross sections are too small.

\begin{figure}[t!] 
\hspace{-0.7cm}
\includegraphics[width=8.cm]{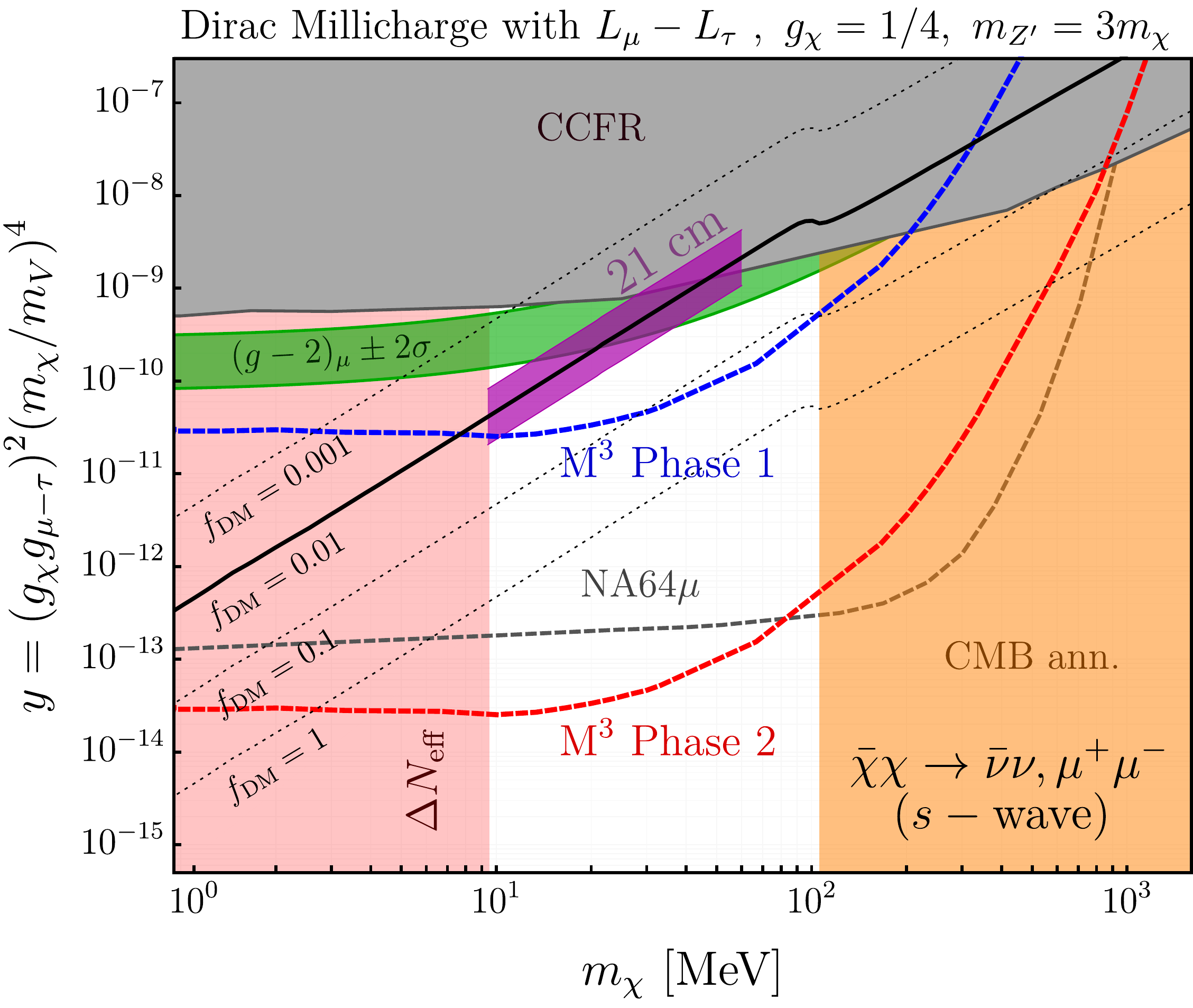}~~~
\includegraphics[width=8.cm]{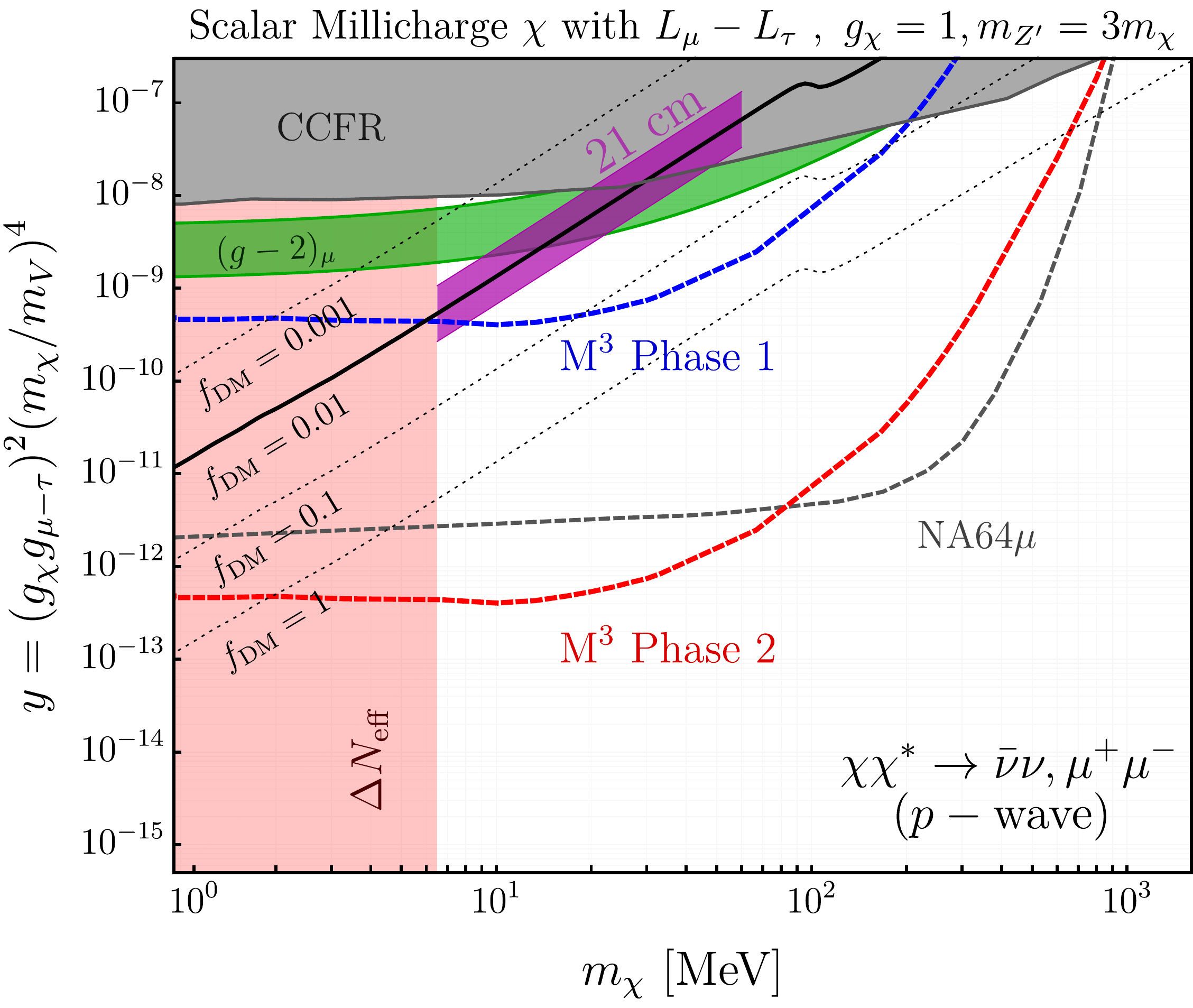} 
   \caption{\label{fig:21cm} Parameter space and projections for fermion (\emph{left}) and scalar (\emph{right}) millicharge particles which constitute a subcomponent of the 
  total dark matter abundance. If such particles also carry appreciable charge under $U(1)_{L_\mu - L_\tau}$,
  they can achieve their subdominant relic abundance via
  $\chi \chi \to \nu \nu, \mu\mu$ annihilation mediated by a virtual $s$-channel $\zp$.
  The black contours represent different fractional DM abundances $f_{\rm DM}$ and the purple region represents the parameter space
  for which such a DM subcomponent can resolve the $\sim 3.8 \sigma$ anomaly in the EDGES 21 cm absorption signal \cite{Berlin:2018sjs}.
    Also plotted are CCFR constraints from the neutrino trident process \cite{Mishra:1991bv,Altmannshofer:2014pba}, $\Delta N_{\rm eff.}$  bounds on
    light DM species \cite{Boehm:2013jpa,Nollett:2014lwa},  CMB annihilation bounds for the Dirac fermion scenario with $s$-wave annihilation \cite{Ade:2015xua},
    and projected sensitivity for the NA64 experiment \cite{Gninenko:2016kpg}. For more details about the DM interpretation of the 21 cm signal, see \cite{Berlin:2018sjs}.}
\vspace{0cm}
\end{figure}

Intriguingly, we also find that both Phase 1 and Phase 2 have sensitivity to a class of DM explanations for the $\sim 3.8 \sigma$ anomaly reported by
the EDGES collaboration \cite{Bowman:2018yin}. It has been shown that a $\sim 1 \% $ subcomponent of DM with a QED millicharge of order $\sim 10^{-3} e$ can cool
the SM gas temperature at redshift $z \sim 20$ and thereby account for the magnitude of the observed absorption feature \cite{Barkana:2018lgd}. However, Ref. \cite{Berlin:2018sjs} pointed 
out  that such a scenario generically requires dark forces to deplete the millicharge abundance 
in the early universe to account for the $\sim 1\% $ fraction needed to resolve the anomaly. One viable possibility for generating the appropriate
abundance is that the millicharges also carry additional $U(1)_{L_\mu - L_\tau}$ charge, which allows $\chi \chi \to \nu \nu, \mu \mu$ annihilation 
processes to set the appropriate DM fraction at late times. In Fig. \ref{fig:21cm} we present the viable 
parameter space for this scenario along with various constraints. The solid and dotted black curves are contours of constant $\chi$ abundance, and the purple region labeled ``EDGES'' represents the favored parameter space for appreciable hydrogen gas cooling 
at $z = 20$ if DM also carries a $\sim 10^{-3}e$ millicharge (see \cite{Berlin:2018sjs} for more details). 

\section{Conclusion}
\label{sec:conclusion}

Light, weakly-coupled particles with muon-philic couplings arise in various extensions of the Standard Model, yet much of their
best-motivated parameter space remains experimentally elusive. Of particular interest are scenarios in which such particles resolve the 
persistent  $(g-2)_\mu$ anomaly and/or serve as mediators that couple to both dark and visible matter. In this paper we have proposed a new muon-beam missing-momentum experiment, M$^3$, to greatly 
improve experimental coverage of invisibly-decaying, muon-philic forces with existing beams and detector technology. 

 Our setup consists of a relativistic $\sim 15$ GeV muon beam impinging on a fixed,
   active target with downstream electromagnetic and hadronic calorimeters to veto  SM particles that emerge 
   undetected from the target. Upstream and downstream of the target, tracking layers measure the incident and outgoing muon momenta, with
  the signal region consisting of evens for which the incident muon loses an order-one fraction of its initial momentum with no 
  detectable energy deposited in the target or veto system. A test phase of our proposal with $10^{10}$ muons on target can exhaustively test all scenarios in which 
   light, invisibly-decaying muon-philic particles resolve the $(g-2)_\mu$ anomaly. A mature realization of our concept with $ 10^{13}$ muons on target can test 
  nearly all predictive sub-GeV dark matter scenarios in which muon-philic forces are responsible for thermal freeze-out. Both phases correspond to presently-feasible
experimental options currently available at Fermilab. 
  
A moderate-energy muon beam missing-momentum experiment has several appealing features. First, the cross sections for most background processes (due to SM particles produced, but not observed) are typically smaller than those of an electron-beam experiment with identical target thickness because most such events are initiated by photonuclear interactions with beam-initiated bremsstrahlung radiation. Since the bremsstrahlung rate for a muon
beam is reduced by a factor of $ (m_e/m_\mu)^2 \sim 2 \times 10^{-5}$ compared to electron beams, our proposal can be implemented with a much thicker target than LDMX, which increases the effective luminosity of the experiment and compensates for the overall reduction of available muons (relative to an ideally-executed electron
beam effort). Second, by operating at $\sim 15$ GeV, a tenth of the nominal muon beam energy of NA64 at the CERN SPS, our proposal allows for much greater beam deflection for recoiling muons as they pass through the tracking layers 
surrounding the active target. This feature allows the experiment to operate on a much smaller ($\sim$ few m) length scale. Furthermore, lower beam energies require less powerful magnetic fields 
to achieve comparable muon angular deflections and hence comparable momentum resolution. 

In this paper, we have primarily motivated our new technique by considering the experimental reach for
 invisibly decaying muon-philic particles produced in muon-nucleus collisions inside the target. 
Indeed, a compelling interpretation of this signature is that
 the new particles decay to dark matter, thereby enabling the technique to probe the couplings that 
 are responsible for setting the relic abundance via thermal freeze-out. However, our case studies 
do not exhaust the full physics search potential of the proposed setup (and possible variations thereof). 
 In particular, it would be interesting to study the possibility of cascade signatures or displaced vertices that involve 
 combinations of missing and visible energy. Such processes are qualitatively distinct from background processes
 initiated in the target and are not covered by the analysis presented here. Furthermore, our 
 setup may also be useful for better understanding rare muon-nucleus scattering processes, which
 could help constrain neutrino-nucleus interactions of relevance for the emerging long baseline 
 neutrino oscillation program. We leave these (and other) questions for future work.

\vspace{3mm}
\noindent {\bf Acknowledgments:}  We thank Chien-Yi Chen, Ross Corliss, Corrado Gatto, Richard Milner, Maxim Pospelov, Philip Schuster, Natalia Toro, Chris Tully, Yimin Wang, and Yiming Zhong for helpful conversations. We especially thank Jerry Annala, Mary Convery, Jim Morgan, Sam Posen, Mandy Rominsky, Diktys Stratakis, and Adam Watts for discussions regarding the Fermilab accelerator complex and test beam.   We also thank the developers of the \texttt{LDMX-sw} code for simulations used in this study.  This work was initiated and performed in part at the Aspen Center for Physics, which is supported by National Science Foundation grant PHY-1607611. Fermilab is operated by Fermi Research Alliance, LLC, under Contract No. DE-AC02-07CH11359 with the US Department of Energy.

\appendix
\section{Relic Density Calculation}
\label{sec:appendix-density}

\renewcommand{\theequation}{A-\arabic{equation}}
\setcounter{equation}{0}
\renewcommand{\thefigure}{A-\arabic{figure}}
\setcounter{figure}{0}

\label{app:relic-density}
To compute the relic density curves presented in Figs. \ref{fig:thermal-fig} and \ref{fig:21cm} in the context
of a gauged $L_\mu -L_\tau$ model, we follow the  thermal freeze-out prescription in \cite{Kolb:1990vq}. For a DM particle $\chi$ with $\lambda_\chi$ internal degrees of freedom, the thermally averaged annihilation rate can be parametrized as $\langle \sigma |v|\rangle  \equiv \sigma_0 x^{-n}$ where $\sigma_0$ is a constant, $x \equiv m_\chi/T$, and $n = 0$ and $1$ for $s$ and $p$-wave annihilation, respectively. The asymptotic $\chi$  abundance at late times is given by
\be\label{eq:density}
\Omega_\chi h^2 = 1.07 \, \times 10^9 \frac{c (n+1) x_f^{n+1} {\rm~ GeV}^{-1}}{(g_{*,S}/g_{*}^{1/2})   m_{Pl} \sigma_0}~~,~~~~~
\ee
where $c = 1$ if the DM is its own antiparticle and $c= 2$ otherwise, $g_{*}$ and $g_{*,S}$ are respectively the effective relativistic and entropic degrees of freedom, $m_{\rm Pl} = 1.22 \times 10^{19} ~\GeV$ is the Planck mass,  and the value of $x$ at freeze-out, $x_f$, satisfies 
\be \label{eq:xf}
 x_f \simeq\log   a  -\left(n+\frac{1}{2}\right) \log  \left( \log a    \right)~~,~~~
 a = 0.038 (n+1)(\lambda_\chi/g_{*}^{1/2})  m_{Pl} m_\chi \sigma_0 ~.~~ 
\ee
 Since $\sigma_0$ cannot be extracted analytically from Eqs. (\ref{eq:density}) and  (\ref{eq:xf}), the system must be solved  numerically. 
The leading order $\chi \bar \chi \to f \bar f$ annihilation cross sections can be straightforwardly extracted using 
the procedure in \cite{Wells:1994qy}. In the representative models we consider in this paper,  $\chi \chi \to \bar f f$ annihilation where $f$ is a charged SM  fermion, we have  
\be \label{eq:sigma0}
~~~~~~\sigma_0 =   \frac{g_\chi^2 g_f^2    (m_f^2 + 2m_\chi^2)}{k \pi \left[  (m_{\zp}^2 -   4 m_\chi^2 )^2    + m_\zp^2 \Gamma_{\zp}^2    \right] } \sqrt{1 - \frac{m_f^2}{m_\chi^2}} ,
\ee 
where $g_{\chi, f}$ is the mediator's coupling to $\chi$ or $f$ particles and the parameters $(n, c, k)$ are 
(0, 2, 2) for a (pseudo-) Dirac fermion,  (1, 2, 12) for a complex scalar and (1, 1, 6) for a Majorana fermion. 
For $\chi \chi \to \nu \nu$ annihilation, the $\zp$ neutrino interaction in Eq.~(\ref{eq:Jmutau}) contains an extra insertion of $P_L$, so 
$\sigma_0$ is smaller by a factor of 2. 
In the $m_{\zp} \gg m_\chi \gg m_f$ limit away from the $m_\zp \sim 2m_\chi$ resonance, appropriate for most of the parameter space considered in this paper,
\be
\sigma_0 \propto  \frac{g_\chi^2 g_f^2 m_\chi^2}{m_{\zp}^4} \equiv \frac{y}{m_\chi^2}~~~, ~~~ y \equiv (g_\chi g_f)^2 \left( \frac{ m_\chi }{ m_{\zp} } \right)^4,
\ee
so for each value of  $m_\chi$, there is a unique value of the dimensionless parameter $y$  which yields thermal freeze-out.  The definition of $y$ used
in this context is adapted from the notation used in \cite{Izaguirre:2015yja,Krnjaic:2015mbs}. Note that in our numerical results, the appropriate $\sigma_0$ includes 
all available annihilation channels, whereas Eq.~(\ref{eq:sigma0}) represents only one possible final state.
Note also that for the gauged $L_{\mu}- L_{\tau}$ results in Fig. \ref{fig:thermal-fig}, we take the SM fermions to carry unit charge and identify $g_f  \to g_{\mu-\tau}$, but allow $g_\chi \ne g_{\mu - \tau}$. Because the DM is vectorlike
under the gauge group, all charge assignments are consistent with anomaly cancellation with no additional matter required.

\section{Reducible Trident Background}
\label{app:Trident}

\renewcommand{\theequation}{B-\arabic{equation}}
\setcounter{equation}{0}
\renewcommand{\thefigure}{B-\arabic{figure}}
\setcounter{figure}{0}

 \begin{figure}[t!] 
\hspace{-0.5cm}
\includegraphics[width=7.5cm]{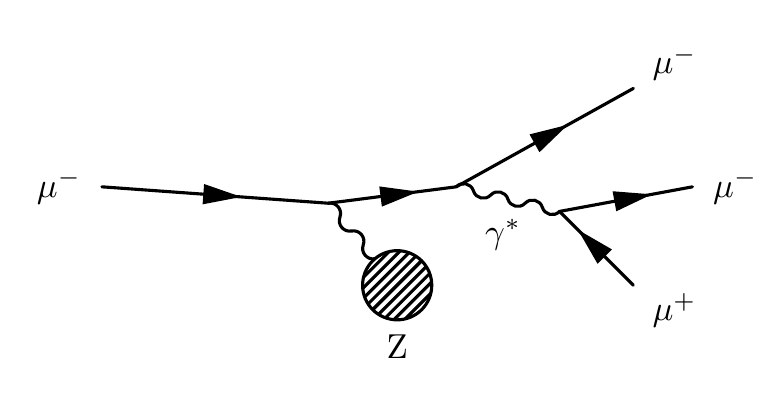}~~
\includegraphics[width=7.5cm]{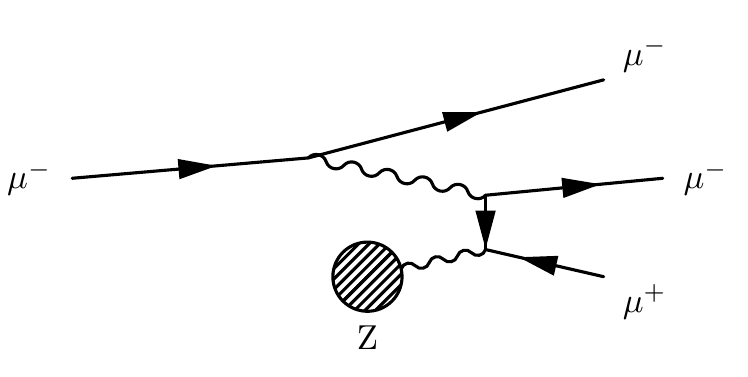}
\caption{Feynman diagrams for muon pair creation: radiative trident (\emph{left}) and Bethe-Heitler trident (\emph{right}). If 
this process takes place inside the target, it can mimic the signal if the opposite-sign muon is not detected and the same-sign pair are collinear with similar momenta (and thus not resolved 
as separate particles based on deflection). 
\label{fig:trident}}
\vspace{0cm}
\end{figure}

A potential reducible background is muon pair creation. Two diagrams contribute to this process, referred to as the radiative trident process and the Bethe-Heitler trident process, shown on the left and right of  Fig.~\ref{fig:trident}, respectively. As discussed in detail in Ref.~\cite{Essig:2010xa}, the Bethe-Heitler trident dominates over the radiative trident, and moreover has a forward singularity where one muon is energetic and collinear with the recoiling muon, and the second muon is soft. This can potentially mimic the signal of a single outgoing muon with significant energy loss if the collinear muon is the same sign as the recoiling muon and has nearly identical momentum, thus producing only a single track to within the angular and spatial resolution of the tracker, and the second muon is soft and/or wide-angle. Indeed, Ref.~\cite{Gninenko:2014pea} finds that the trident is a dominant background in their proposal.

\begin{figure}[t!] 
\hspace{-0.5cm}
\includegraphics[width=7.6cm]{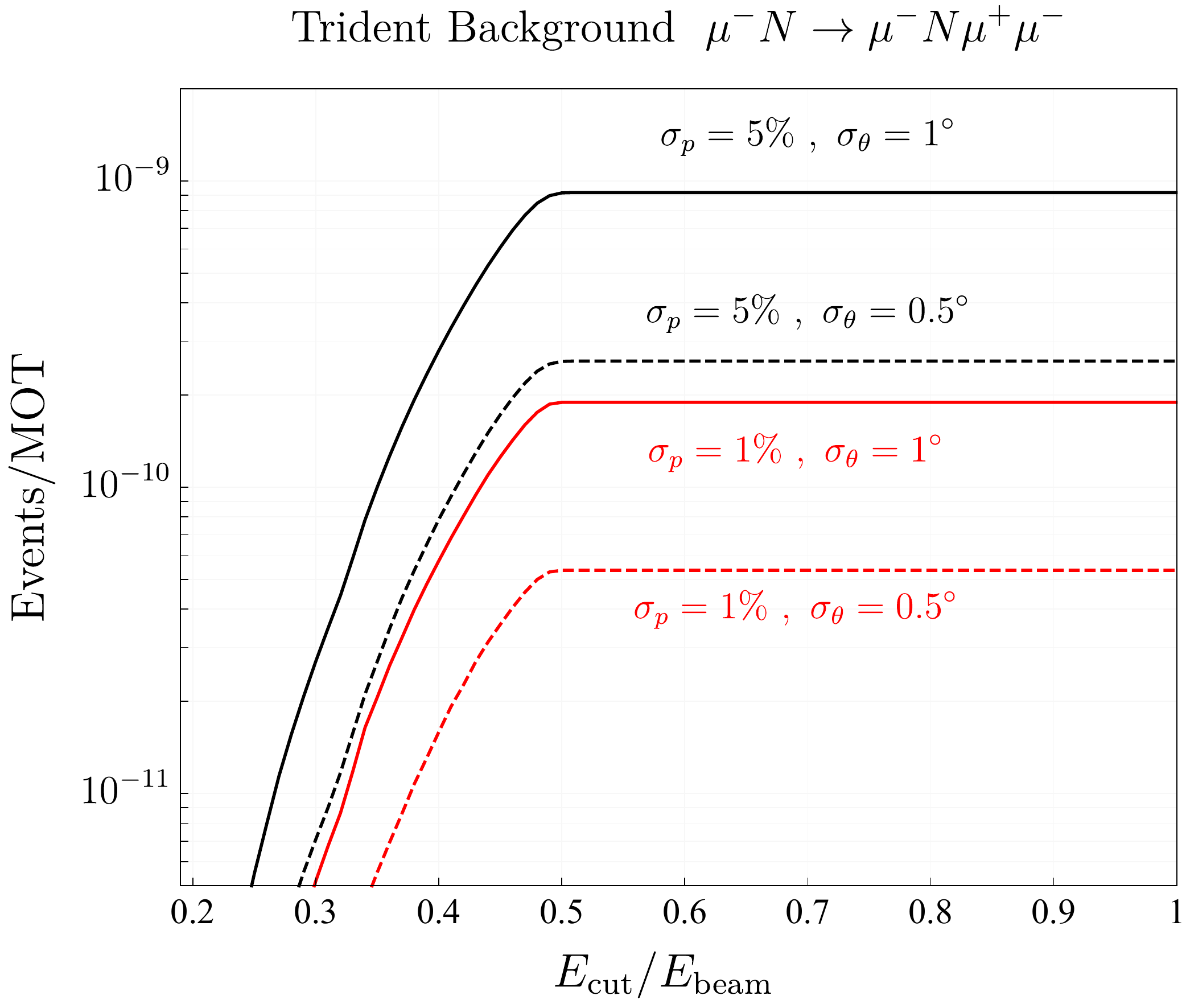}~~~
\includegraphics[width=7.6cm]{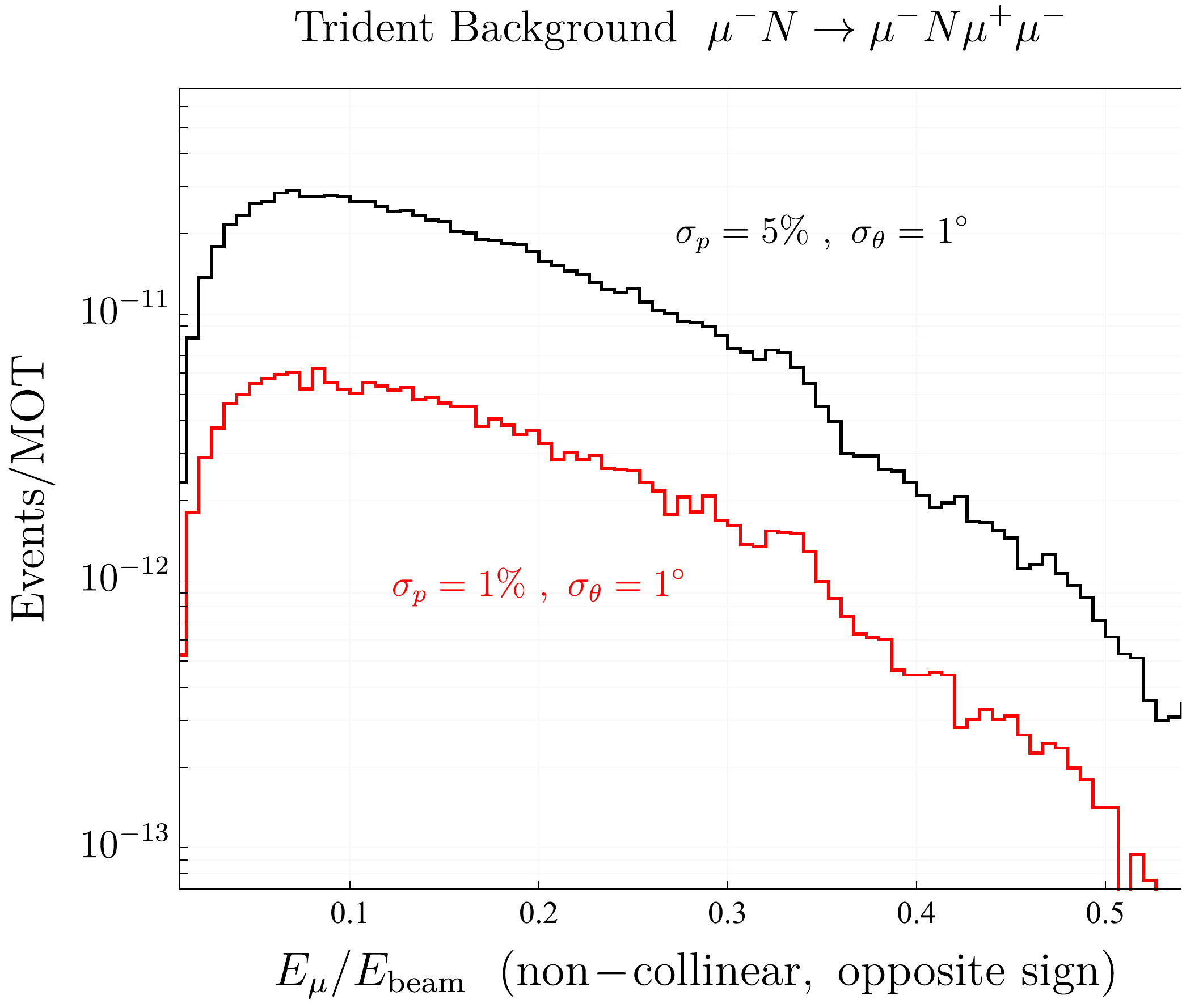}
\caption{(\emph{Left}) Total muon trident background events $\mu N \to \mu N \mu \mu $ for which 
the two same-sign muons are collinear to within various energy $\sigma_E$ and 
angular $\sigma_\theta$ resolutions plotted
as a function of the signal energy threshold $E_{\rm cut}$.
(\emph{Right}) Differential energy spectrum for the opposite
sign, non-collinear muon for $E_{\rm cut} = 8$ GeV,  $\sigma_\theta= 1^\circ $ and
different choices of $\sigma_p$. 
\label{fig:tridentspectrum}}
\vspace{0cm}
\end{figure}

Following the method of Ref.~\cite{Essig:2010xa}, we simulated the Bethe-Heitler diagram only (ignoring interference effects) in \texttt{MadGraph}, selecting events for which two same-sign muons were collinear to within angular resolution $\sigma_\theta$ and the momenta were matched to resolution $\sigma_p$. Fig.~\ref{fig:tridentspectrum} (left) shows the event rate for this potential background as a function of the signal energy threshold for various choices of the energy and angular resolutions. The background rate saturates above a cut at 50\% of the beam energy because if the two collinear muons have nearly identical energy, each must carry less than half the beam energy, thus trivially satisfying the cut. For fairly conservative resolutions $\sigma_p = 5\%$, $\sigma_\theta = 0.5^\circ$, the maximum trident rate is about 3 events for $10^{10}$ MOT. Moreover, if the third muon is not too soft or wide-angle, it can be detected in the tracker and vetoed. Fig.~\ref{fig:tridentspectrum} (right) shows the energy spectrum of the non-collinear muon.

Crucially, our proposal has another strong handle on vetoing the trident background. Even if the two same-sign muon tracks are exactly parallel and have exactly the same momentum, both muons are minimum-ionizing particles, and thus would deposit twice the expected energy in the tracker as would a single particle. In each layer of the HCAL, $\mathcal{O}(10)$ photoelectrons will be produced per MIP per layer. With 10 layers, a single muon would yield $\mathcal{O}(100)$ photoelectrons while two parallel muons would yield $\mathcal{O}(200)$; assuming Poisson statistics, this results in a fake rate of less than $10^{-4}$. The HCAL will likely have many more than 10 layers, but this is already sufficient to bring the Bethe-Heitler rate below $10^{-13}$ per MOT, rendering this background negligible for both Phase 1 and Phase 2.

\section{Irreducible CCQE Background}
\label{app:CCQE}

\renewcommand{\theequation}{C-\arabic{equation}}
\setcounter{equation}{0}
\renewcommand{\thefigure}{C-\arabic{figure}}
\setcounter{figure}{0}
The signal considered in this paper is $\mu(E_{\rm beam}) N \to \mu(E \ll E_{\rm beam}) N (S,V)$, followed
by an invisible $S$ or $V$ decay.  
The main irreducible background for this process is $\mu e \to \mu e$ in which the recoiling electron receives 
a large fraction of the incident muon energy and undergoes an $e p \to n \nu$ CCQE conversion before it can be vetoed as an electromagnetic
shower inside the active target.

The differential cross section for fixed target $\mu e \to \mu e$  scattering is 
\be
\label{eq:dsigma}
\frac{d\sigma}{dE_e} = \frac{2 \pi \alpha^2}{ m_e  E_e^2} \left( 1 - \frac{E_e}{E_\mu} +  \frac{E^2_e}{2 E^2_\mu}  \right)   ,
\ee
where $E_\mu$ is the incident muon beam energy, $E_e$ is electron recoil energy, and both are evaluated in the lab frame. 
The probability of a CCQE background event with $e$ recoil energies above threshold $E_{\rm cut}$ in a material of electron density $n_e$ and infinitesimal thickness $dz$ is 
\be
\label{eq:dP}
dP_{ \mu e \to \mu e \rm + CCQE} =  dz \, n_e \int_{E_{\rm cut}}^{E_\mu} dE_e \frac{d\sigma}{dE_e}   P_{\rm CCQE}   \,  \Theta \!\left( E_\mu - E^{\rm min}_\mu\right) ,
\ee
where the $\Theta$ function enforces energy conservation,  
\be
E^{\rm min}_\mu = \frac{E_e}{2} \left[ 1 + \sqrt{ \left(1 + \frac{2 m_e}{E_e}\right)\left( 1 + \frac{2 m_\mu^2 }{m_e E_e} \right)   \! \!  } \>\> \right]
\ee
is the minimum beam energy required to induce an electron recoil with energy $E_e$, and 
\be
P_{\rm CCQE} =  n_p \sigma_F (\ell - z) ~~,~~ \sigma_F = \frac{G_F^2 m_p E_e}{2 \pi}~~,
\ee
is the probability of a CCQE conversion event for an electron with energy $E_e$ passing through a target with proton density $n_p = n_e$ and 
longitudinal length $(\ell-z)$, corresponding to the amount of material an electron encounters after forward scattering at position $z \in (0, \ell)$ 
in the one-dimensional approximation. Keeping only the leading term of Eq.~(\ref{eq:dsigma}) and integrating over $z$ in Eq.~(\ref{eq:dP}), the
 probability of a CCQE event per incident muon in a tungsten target is 
\be
P_{ \mu e \to \mu e \rm + CCQE} \approx   ( \alpha\, n_e  G_F    \ell)^2  \frac{m_p}{2 m_e}  \log \frac{ E_{ \mu} }{ E_{\rm cut}} \simeq 5 \times 10^{-15},
\ee
where, in the last step we have taken $\ell = 50 X_0 = 17.5$ cm, $n_e = 4.6 \times 10^{24} \, \cm^{-3}$, and $E_{\rm cut} =  E_{\rm \mu}/2$. 
This estimate is conservative because we have neglected the kinematic restriction imposed by the $\Theta$ function in Eq.~(\ref{eq:dP}), which
reduces the overall number of scattering events; evaluating the full expression gives $P_{ \mu e \to \mu e \rm + CCQE} = 3 \times 10^{-16}$ for the 
same setup. 

Furthermore, unlike LDMX with an electron beam, which requires a thin, passive target, using a muon beam with a thicker, active target  makes it possible
to veto events with electromagnetic showers inside the much larger target. As such, the above estimate is overly conservative by integrating the contribution 
over the total target length; in practice this process could only happen for lengths of order $ \sim X_0$ without being vetoed. 
Thus, we conclude that the CCQE background is not relevant even for the high luminosity Phase 2 run with a benchmark luminosity of $10^{13}$ MOT.

\bibliography{MuonProtonBib.bib}

\end{document}